\documentclass[10pt]{article} 
\usepackage[preprint]{tmlr}

\usepackage{amsmath,amsfonts,bm}

\def\eqref#1{equation~\ref{#1}}

\def\1{\bm{1}}

\DeclareMathAlphabet{\mathsfit}{\encodingdefault}{\sfdefault}{m}{sl}
\SetMathAlphabet{\mathsfit}{bold}{\encodingdefault}{\sfdefault}{bx}{n}

\usepackage{hyperref}
\usepackage{url}
\usepackage{booktabs}
\usepackage{multirow}
\usepackage{graphicx}
\usepackage{caption}
\usepackage{subcaption}

\definecolor{ruddy}{rgb}{1.0, 0.0, 0.16}
\definecolor{gblue}{RGB}{29, 144, 255}
\definecolor{royalblue}{rgb}{0.25, 0.41, 0.88}

\hypersetup{
  colorlinks,
  citecolor=royalblue,
  linkcolor=ruddy,
  urlcolor=royalblue}

\title{\raisebox{3pt}{\LARGE A Survey on Responsible Generative AI: What to Generate} \\  and What Not}

\author{\name Jindong Gu \\ 
\addr University of Oxford, United Kingdom \\
}

\def\month{MM}  
\def\year{YYYY} 
\def\openreview{\url{https://openreview.net/forum?id=XXXX}} 

\begin{document}

\maketitle

\vspace{-0.45cm}

\begin{abstract}
In recent years, generative AI (GenAI), like large language models and text-to-image models, has received significant attention across various domains. However, ensuring the responsible generation of content by these models is crucial for their real-world applicability. This raises an interesting question: \textit{What should responsible GenAI generate, and what should it not?} To answer the question, this paper investigates the practical responsible requirements of both textual and visual generative models, outlining five key considerations: generating truthful content, avoiding toxic content, refusing harmful instruction, leaking no training data-related content, and ensuring generated content identifiable. Specifically, we review recent advancements and challenges in addressing these requirements. Besides, we discuss and emphasize the importance of responsible GenAI across healthcare, education, finance, and artificial general intelligence domains. Through a unified perspective on both textual and visual generative models, this paper aims to provide insights into practical safety-related issues and further benefit the community in building responsible GenAI.
\end{abstract}

{
  \hypersetup{linkcolor=black}
  \tableofcontents
}

\section{Introduction}\label{Sec:Introduction}

Generative AI (GenAI) has received remarkable attention recently. Various generative models have been developed in diverse domains, for example, autoregressive large language models~\citep{brown2020language,touvron2023llama} and text-to-image generative models~\citep{saharia2022photorealistic,rombach2022high,betker2023improving}. In real-world applications, the generated contents have to be not only high-quality but also responsible. Thus, this raises the question: What should responsible GenAI generates, and what not?

In this paper, we summarize the responsible requirements of current generative models. Two types of generative models are mainly considered in this work, namely, textual generative models and visual ones. Specifically, textual generative models generate textual responses based on textual or visual inputs, which include autoregressive large language models~\citep{brown2020language,touvron2023llama} and multimodal LLMs~\citep{geminiteam2023gemini,openai2023gpt4}. Similarly, visual generative models are the models that generate images or videos based on textual and visual inputs~\citep{rombach2022high,betker2023improving,sora}.

For both types of generative models, we summarize five responsible requirements for the generated contents. The requirements are not comprehensive, we only review five popular requirements that are important to both academic and industrial communities. Concretely, the five requirements are as follows:

\textbf{1. To generate truthful content.} The generated content is expected to be truthful. However, current generative models could generate content that strays from factual reality or includes fabricated information, which is known as \textit{Hallucination}~\citep{maynez2020faithfulness}. For instance, language models would generate non-fact, and text-to-image models create images with objects that are not specified in the text prompts. Many efforts have been made to identify, elucidate, and tackle the problem of hallucination~\citep{huang2023survey}.

\textbf{2. Not to generate toxic content.} It is well known that both language models and image generation models could generate biased content~\citep{sap2019social,naik2023social}. More recently, with wide applications, it has been found that toxic content could also be generated as responses to end-users. The goal to make generated content unbiased and non-toxic has been intensively explored from various perspectives, e.g., filtering training data and fine-tuning~\citep{ganguli2022red,friedrich2023fair}.

\textbf{3. Not to generate content for harmful instructions.} With or without safety alignment, the current generative AI model can still generate inappropriate content given the adversarial prompts, which is known as \textit{Jailbreak}~\citep{perez2022ignore}. For instance, the textual generative models would output the details for the input prompt 'How to build bombs?' + adversarial prompt. Visual generative models would generate inappropriate images when adversarial text prompts are given~\citep{yang2024sneakyprompt}. Much effort has been made to reveal such vulnerability and defend against these adversarial prompts.

\textbf{4. Not to generate training data-related content.} Recent generative models are often large-scale and have a large number of parameters. Recent research shows that the learned parameters contain the information of training instances. For instance, training text from a language model can be extracted~\citep{carlini2021extracting}, and training images can be synthesized with the corresponding pre-trained text-to-image model~\citep{carlini2023extracting}. How to better extract training data from pre-trained generative models and how to hide the training data information have been intensively studied in our community.

\textbf{5. To Generate identifiable content.} The copyright of generated content is a complicated problem, which requires the knowledge of multiple disciplines. There are multiple levels of copyright problems. One is how to generate detectable content, e.g. with watermarks~\citep{usop2021review}. To attribute the copyright, another intensively studied topic is model attribution~\citep{uchendu2020authorship}. It aims to identify which generative model generates a particular instance.

Our paper is organized as follows: we recall the preliminary knowledge of GenAI and the vulnerability of deep neural networks in Sec.~\ref{sec:pre}. Sec.~\ref{sec:txtGenAI} summarizes the research of textual generative models on responsible generation, while Sec.~\ref{sec:t2iGenAI} presents the research about responsible visual generation. Besides, we also discuss the application of GenAI in different domains from the perspective of responsible AI in Sec.~\ref{sec:apps}. The involved domains include healthcare, education, finance, and artificial general intelligence. Furthermore, we discuss the challenges and opportunities of responsible GenAI in Sec.~\ref{sec:dis} and conclude our paper in Sec.~\ref{sec:conclu}.

Our paper differs from related works in the following three points: 1) We present the research topics on responsible GenAI and their recent progress. Especially, we provide a unified perspective for both textual generative models and visual generative models. 2) We summarize the practical safety-related problems that both academics and industry have intensively worked on. 3) We discuss the risks and concerns when applying GenAI to various domains, including general-purpose intelligent systems.

\section{Preliminaries}
\label{sec:pre}
In this section, we provide background knowledge on the techniques used to build safe generative models, reveal their vulnerabilities, and defend against malicious inputs. Specifically, we begin by reviewing the foundational components of modern generative AI. Then, we delve into the fundamentals of adversarial attacks and backdoor attacks on deep neural networks.

\subsection{Preliminary of Modern Generative AI}
For textual and visual generative models, we first introduce the popular model architectures (i.e., Transformer and Diffusion Models) and then present basic knowledge of pre-training and post-training, respectively. 

\subsubsection{Transformer-based Textual Generative AI}
\label{sec:text_pre}
\noindent\textbf{Transformer Architecture.} Transformer~\citep{vaswani2017attention} is often composed of a list of self-attention blocks consisting self-attention layer, MLP layer as well as other operations. The self-attention layer is the main component of the self-attention block, which takes a sequence of tokens and outputs a sequence of tokens with new embeddings. Concretely, they can be expressed as follows. Given the input consists of a list of tokens $\pmb{x} \in \mathbb{R}^{(N \times M)} $, the queries, keys, and values of them are computed as
\begin{equation}
\pmb{K} = \pmb{W}_k \cdot \pmb{x}, \;\;\;\; 
\pmb{Q} = \pmb{W}_q \cdot \pmb{x}, \;\;\;\;
\pmb{V} = \pmb{W}_v \cdot \pmb{x},   
\end{equation}
where $\pmb{W}_k\in \mathbb{R}^{(M \times D)}, \pmb{W}_q\in \mathbb{R}^{(M \times D)}$ and $\pmb{W}_v\in \mathbb{R}^{(M \times D)}$ are linear mapping matrix. 

The attention between the input tokens is
\begin{equation}
    \pmb{A} =\; \mathrm{Softmax}(\pmb{Q}\cdot(\pmb{K})^T/\sqrt{D}).
\end{equation}
The output embeddings $\pmb{Z}$ of the self-attention layer are
\begin{equation}
\pmb{Z} = \pmb{A} \cdot \pmb{V}.
\label{equ:v}
\end{equation}
Transformer-based encoder-decoder~\citep{raffel2020exploring}, decoder-only architectures~\citep{openai2023gpt4} are often applied as generative models, while encoder-only architectures~\citep{devlin2018bert} are designed for discriminative tasks. We now introduce the most popular architecture, namely, decoder-only architecture with a masked self-attention layer. Different from the self-attention, the masked self-attention replaces the Equation~\ref{equ:v} with the following equation:
\begin{equation}
    \pmb{Z} = (\pmb{A} + \pmb{M}) \cdot \pmb{V},
\end{equation}
where $\pmb{M} \in \mathbb{R}^{(D \times D)}$ is defined as  
\begin{equation}
    \pmb{M}(i, j) = 
    \begin{cases}
    0, & \text{if } i \geq j\\
    -\infty, & \text{if } i < j
    \end{cases},
\end{equation}
where $i$ and $j$ are index of the dimension of token embedding $D$.

The architecture above is often applied for autoregressive generation. Specifically, given N input tokens, a generative model generates the $N+1$-th token. The predicted token is appended to the previous N tokens. The model generates the $N+2$-th token based on the $N+1$ tokens. The probability of generating $y_{N+2}$ as the $N+2$-th token is
\begin{equation}
    \mathbb{P}(y_{N+2}\; |\; y_1,\; y_2,\; \cdots,\; y_{N+1}),
\end{equation}
The autoregressive generation stops until a certain token is generated. Multi-head of self-attention mechanism can be computed in parallel. Their outputs are concatenated as the final token embedding.

\textbf{Pre-training of Decoder-only Transformer.} We first introduce the data preparation for autoregressive pre-training as the following steps: 
\begin{enumerate}
    \item Data Collection and Cleaning: A large and diverse dataset of text is collected from various sources such as books, articles, websites, forums, and other textual repositories. The collected data is then cleaned, e.g., by removing irrelevant characters and handling special cases like punctuation.
    \item Sequence Creation: In this step, the raw text data is divided into sequences of fixed or variable length. Each sequence represents a contiguous segment of text, and these sequences serve as the basic units of input for the model pretraining.
    \item Tokenization: After the sequences are created, each sequence is tokenized into individual tokens. Tokenization involves splitting the text into smaller units such as words, subwords, or characters, depending on the tokenization scheme chosen for the model. This step converts the text into a sequence of discrete tokens.
    \item Special Tokens Addition: Special tokens may be added to the sequences to indicate the beginning and end of each sequence, as well as to mark padding and unknown tokens. 
\end{enumerate}
For each input sequence consisting of a list of tokens $\pmb{x} = \{x_1, x_2, \cdots, x_N\}$, the joint distribution is computed as follows
\begin{equation}
    \mathbb{P}(\pmb{x}) = \mathbb{P}({x_1, x_2, \cdots, x_N}) = \prod_{i=1}^N \mathbb{P}(x_i \mid x_1, x_2, \cdots, x_{i-1}),
\end{equation}
The goal is to maximize the joint probability by updating the parameters of the autoregressive model.

\textbf{Post-training of Transformer-based Language Models}
LLM pre-training on large-scale datasets does not inherently make them capable of following users' instructions. The output of LLMs can be not helpful or even harmful to the users. To ensure that LLMs generate useful and responsible responses, post-training is often conducted to align them with human intent. Various post-training strategies have been proposed. They often start with Supervised Fine-Tuning (SFT). For SFT, a dataset is first collected where the labelers provide demonstrations of the desired behavior on the input prompts. The pre-trained LLM is fine-tuned on the collected dataset following a standard setting. Specifically, SFT is similar to the pre-training process where training samples are constructed by concatenating a prompt $\{x_1, x_2, \cdots, x_N\}$ and a desired response $\{x_{N+1}, x_{N+2}, \cdots, x_{N+L}\}$.
\begin{equation}
\label{eq:sft}
    \mathbb{P}({x_{N+1}, \cdots, x_{N+L} \mid x_1, x_2, \cdots, x_N}) = \prod_{i=N+1}^{N+L} \mathbb{P}(x_i \mid x_1, x_2, \cdots, x_N, x_{N+1}, \cdots, x_{N+L}), 
\end{equation}

The SFT model generates task-specific completions. However, its responses may violate safety rules. To address it, Reinforcement Learning with Human Feedback (RLHF)~\citep{christiano2017deep,ouyang2022training} is applied to integrate human preference, in which the human preference is first modeled as a reward in Reward modeling (RM), and Reinforcement Learning (RL) is applied to maximize the reward model via updating the model. The two steps are introduced below.

1) Reward modeling. For RM, a dataset of comparisons between model outputs is collected, where labelers rank the model outputs for each given input. Concretely, K model responses are sampled for each input. Any two of them are ranked by the labelers, namely, there are $K \choose 2$ annotations for each input prompt.

A model, dubbed reward model, is trained on the collected dataset $D$. The reward model $r_{\theta}(\cdot)$, parameterized by $\theta$, takes in a prompt and the model response on it and outputs a scalar reward. The model is expected to output a larger scale value for the preferred response than that for the other response. Specifically, the loss function for the reward model can be formulated as follows: 
\begin{equation}
L(\theta)=\mathbb{E}_{(\pmb{x}, y_{w}, y_{l}) \sim D}[\log (\sigma(r_{\theta}(\pmb{x}, y_{w})-r_{\theta}(x, y_{l})))]
\label{equ:reward}
\end{equation}
where $r_{\theta}(\pmb{x}, y)$ is the scalar output of the reward model when the prompt $\pmb{x}$ and the model response $y$ on it are taken as input, $y_{w}$ is the preferred response out of the pair of $y_{w}$ and $y_{l}$. Note that all ${K \choose 2}$ comparisons from each prompt are taken as a single batch to avoid overfitting~\citep{ouyang2022training}.

Finally, the scalar outputs of the reward model are normalized so that the demonstrations provided by labelers achieve a mean score of 0 before tuning the model with RLHF. This is meaningful because the RM training loss defined as in Equation~\ref{equ:reward} is invariant to shifts in reward. More details regarding reward modeling can be found in~\citet{ouyang2022training}. Further research shows that LLM can exploit errors in the not-perfect reward model to achieve a high reward, which is dubbed reward hacking. Efforts have been made to mitigate the hacking phenomenon~\citep{amodei2016concrete,coste2023reward,eisenstein2023helping}.

2) Reinforcement learning. When a reward model is available, SFT model is fine-tuned with RL to maximize the rewards received by the reward model. Given the prompt and the model's reponse to it, a reward is returned by the reward model, which is used to update model parameters. In addition, a per-token KL penalty from the SFT model at each token is applied to mitigate over-optimization of the reward model~\citep{ouyang2022training}.

The model to be tuned can be seen as a policy network $\pi_{\phi}^{\mathrm{RL}}(\cdot)$, parameterized by $\phi$. Given a prompt (i.e., part of the environment), the model makes a response based on the prompt, which can be seen as an action. The output of the reward model for the prompt and the response is the reward returned by the environment. Then, the episode ends. The model is updated to maximize the following objective:
\begin{equation}
L(\phi)= \mathbb{E}_{(\pmb{x}, y) \sim D_{\pi_{\phi}^{\mathrm{RL}}}}[r_{\theta}(\pmb{x}, y)-\beta \log (\pi_{\phi}^{\mathrm{RL}}(y \mid \pmb{x}) / \pi^{\mathrm{SFT}}(y \mid \pmb{x}))] 
\end{equation}
where the first term is the reward, the second one corresponds to the KL penalty for regularization, $\pi_{\phi}^{\mathrm{RL}}$ is the learned policy, \( \pi^{\mathrm{SFT}}\) is the supervised trained model. 

The model tuned with the loss above often shows performance regressions on public NLP datasets. To address it, it is also common to mix the pretraining gradients into the PPO gradients, namely, add the term $\mathbb{E}_{x\sim D_{\mathrm{pretrain}}}\log (\pi_{\phi}^{\mathrm{RL}}(\pmb{x}))$ to the loss objective, where $D_{\mathrm{pretrain}}$ is the distribution of pretrianing datasets~\citep{ouyang2022training}.

In addition to RLHF, many alternative alignment techniques have been proposed to remedy the safety issue, such as controlled decoding~\citep{yang2021fudge,mudgal2023controlled}, sequence likelihood calibration~\citep{zhao2022calibrating}, direct preference optimization~\citep{rafailov2024direct}, and best-of-n finetuning~\citep{touvron2023llama,beirami2024theoretical}. In this preliminary part, we introduce the classic alignment method, i.e., RLHF.

\subsubsection{Diffusion Model-based Visual Generative AI}
\noindent\textbf{Diffusion Model Architecture.}
Diffusion probabilistic models~\citep{sohl2015deep}, also called Diffusion Models (DMs), are designed to fit complex data distribution while keeping tractable. Based on DMs, denoising diffusion probabilistic models (DDPMs)~\citep{ho2020denoising} are proposed for the domain of image generation. The training of DDPM consists of a multi-step forward process and an iterative reverse process. In each step of the forward pass, gaussian noises are added to the natural images. Formally speaking, given a clean image $\pmb{x}_0$ from a distribution $q$, diffusion step $T$ and hyperparameter $\beta_t$, the forward process is following
\begin{equation}
    q(\pmb{x}_{1:T}|\pmb{x}_0) = \prod_{t=1}^T q(\pmb{x}_t \mid \pmb{x}_{t-1}),
\end{equation}

\begin{equation}
    q(\pmb{x}_t \mid \pmb{x}_{t-1})=\mathcal{N}(\pmb{x}_t ; \sqrt{1-\beta_t} \pmb{x}_{t-1}, \beta_t I),
\end{equation}
where $\mathcal{N}(\pmb{x}_t ; \mu, \sigma)$ means sampling gaussian noise with mean of $\mu$ and variance of $\sigma$. The image with added noises at the $t$-th step can be reformulated as 
\begin{equation}
    q(\pmb{x}_t \mid \pmb{x}_0)=\mathcal{N}(\pmb{x}_t ; \sqrt{\bar{\alpha}_t} \pmb{x}_0,(1-\bar{\alpha}_t) I),
\end{equation}
where $\alpha_t= 1-\beta_t, \bar{\alpha_t}= \prod^t_{s=0}\alpha_s$.

Corresponding to the forward process defined above, the reverse process reconstructs images from noisy data $\pmb{x}_T$ iteratively. In the $t$-th iteration of the reverse process, a denoising network predicts the noise that is added to natural image $\pmb{x}_0$ at the $t$-th step of the forward process. The predicted noise is expressed as
$\boldsymbol{\epsilon}_\theta(\pmb{x}_t, t)$
where $\boldsymbol{\epsilon}_\theta(\cdot)$ is the noise prediction network parameterized by $\theta$. Then, the $\pmb{x}_{t-1}$ can be reconstructed from $\pmb{x}_{t}$ and the predicted noise. The noise prediction network is optimized with the following formula
\begin{equation}
    L(\theta)=\mathbb{E}_{t, \mathbf{x}_0, \boldsymbol{\epsilon}}[\|\boldsymbol{\epsilon}-\boldsymbol{\epsilon}_\theta(\sqrt{\bar{\alpha}_t} \pmb{x}_0+\sqrt{1-\bar{\alpha}_t} \boldsymbol{\epsilon}, t)\|^2],
\label{equ:diff_loss}
\end{equation}
where $t$ is uniform between $1$ and $T$, $\boldsymbol{\epsilon} \sim \mathcal{N}(\mathbf{0}, \mathbf{I})$ is random noise. The parameter $\theta$ is updated to minimize the loss defined above.

During inference, DDPM generates images from noisy data within a predefined number of steps, following the reverse process described earlier. However, when the input image dimension is excessively large, scaling DDPMs becomes challenging. To tackle this scalability issue, latent DDPMs have been introduced~\citep{rombach2022high}. These models first map raw images to a lower-dimensional latent space, where the diffusion process is then carried out. The final embedding is mapped to image space again with a decoder.

Besides, condition information has been explored to guide the generated content~\citep{ho2020denoising,rombach2022high}. The condition information $\pmb{c}$ like textual prompts or images are taken as conditional inputs of the noise prediction network in both training and inference processes. With the conditional information, the noise prediction network is optimized as follows:
\begin{equation}
    L_c(\theta)=\mathbb{E}_{t, \mathbf{x}_0, \boldsymbol{\epsilon}}[\|\boldsymbol{\epsilon}-\boldsymbol{\epsilon}_\theta(\sqrt{\bar{\alpha}_t} \pmb{x}_0+\sqrt{1-\bar{\alpha}_t} \boldsymbol{\epsilon},\; t,\; \pmb{c})\|^2],
\label{equ:con_diff_loss}
\end{equation}
If conditional information is integrated into the training process, the generated content can be controlled by user-specified conditional information in inferences.

\textbf{Pre-training of DDPM.} For training unconditional DDPMs~\citep{ho2020denoising}, a dataset comprising images is collected. The parameters in the noise prediction network of DDPM are optimized to minimize a predefined training objective outlined in Equation~\ref{equ:diff_loss}. Conversely, in the training of conditional DDPMs, each training image is typically accompanied by conditional information, such as text, another image, or even audio. For instance, conditional latent diffusion models are trained on a large-scale dataset consisting of image-text pairs~\citep{rombach2022high}. The features of the conditional information are extracted using a feature extractor and fed to the noise prediction network. The optimization objective of the noise prediction is described by Equation~\ref{equ:con_diff_loss}.

\textbf{Post-training of DDPM.} Similar to post-training of LLM, reinforcement learning with human feedback
(RLHF) has also been applied to fine-tune diffusion models~\citep{lee2023aligning,black2023training,fan2024reinforcement,wu2023human,xu2024imagereward}. The construction of a reward model of RLHF requires extensive datasets, optimal architecture, and manual hyperparameter tuning, which makes the process both time and cost-intensive~\citep{yang2023using}. Inspired by Direct Preference Optimization~\citep{rafailov2024direct}, \citet{yang2023using} propose to fine-tune diffusion models directly with Denoising Diffusion Policy Optimization method. In addition to the alignment method above, controllable generation has also been explored to implement the intents of users~\citep{ruiz2023dreambooth,gal2022image,zhang2023adding,liu2022compositional,du2023reduce}.

\subsection{Vulnerability of Deep Neural Networks}
In this section, we provide a preliminary about attacks on deep neural networks, focusing on two main types: adversarial attacks and backdoor attacks. Adversarial attacks seek to fool deep neural networks by altering their inputs during inference, while backdoor attacks aim to induce malicious behaviors in models by interfering with the training process, such as by adding poisoned samples with specific trigger patterns to the training dataset.

\subsubsection{Adversarial Attacks}
\citet{szegedy2013intriguing} find an intriguing property of neural networks that when added to the image, a certain imperceptible perturbation can cause the network to misclassify an image. The adversarial perturbation can be created as follows. Given an input $x$, a model $f(\cdot)$ and its output $f(\pmb{x})$, an adversarial perturbation $\pmb{\delta}$ is created to increase the loss $ \mathcal{L} (f(\pmb{x}+\pmb{\delta}), \pmb{y})$ where $\mathcal{L}(\cdot)$ is the standard cross-entropy loss and $\pmb{\delta}$ is often set to be $\ell_p$-bounded to achieve imperceptibility. The created perturbation corresponding to high loss can mislead the prediction of the model when added to the input.

The optimization of the perturbation can be formulated as follows. The one-step \textit{Fast Gradient Sign Method} (FGSM~\citep{Goodfellow2015ExplainingAH}) creates perturbations as
\begin{equation}
\pmb{\delta} = \epsilon \cdot \text{sign}(\nabla_{\pmb{\delta}} \mathcal{L} (f(\pmb{x}+\pmb{\delta}), \pmb{y})), 
\label{eq: FGSM}
\end{equation}
where $\text{sign}(\cdot)$ is the sign function and $\epsilon$ is a step size corresponding to the allowed perturbation bound.

FGSM with a single-step optimization only achieves limited attack performance. To improve the adversarial effectiveness, the multi-step \textit{Projected Gradient Descent} (PGD~\citep{madry2017towards}) is proposed. Each step of PGD can be expressed as
\begin{equation}
\pmb{\delta}^{t+1} \leftarrow  \text{clip}_{\epsilon} (\pmb{\delta}^{t} + \alpha \cdot  \text{sign}(\nabla_{\pmb{\delta}}  \mathcal{L}  (f(\pmb{x}+\pmb{\delta}), \pmb{y}))),
\label{eq: PGD}
\end{equation}
where $\pmb{\delta}^{t}$ corresponds to the perturbation of $t$-th step and $\text{clip}_{\epsilon}(\cdot)$ is a clipping function that clips its input into $\epsilon$-ball of the input for visual imperceptibility. 

More optimization methods have been proposed to further improve attack effectiveness~\citep{moosavi2016deepfool,carlini2017towards,dong2018boosting,croce2020reliable}. Apart from $\ell_p$-bounded attacks, other attacks with different constraints have also been intensively studied, e.g., sparse attacks~\citep{croce2019sparse,modas2019sparsefool}, patch attacks~\citep{brown2017adversarial,gu2022vision,gu2022evaluating}, semantic attacks~\citep{joshi2019semantic,wang2023semantic}, viewpoint attacks~\citep{dong2022viewfool}, and physical attacks~\citep{huang2020universal,eykholt2018robust}. Furthermore, adversarial attacks on neural networks with text inputs (e.g., language models) have also been explored where the input perturbations are often character-level, word-level, and sentence-level addition, removal, and replacement~\citep{morris2020textattack,zhang2020adversarial}. Different from image space, the text input space is discrete, which poses the main challenges when attacking and defending NLP models~\citep{dinan2019build,sinha2023break}.

An intriguing property of adversarial perturbation is the transferability of adversarial examples, where perturbations crafted for one model can deceive another,
often with a different architecture~\citep{Goodfellow2015ExplainingAH,papernot2016transferability,gu2021effective,ma2023improving,yu2023reliable}. The property poses practical threats to real-world applications since it enables attacks without access to the target model. We refer the reader to the survey paper~\citep{gu2023survey} for more details. To defend against adversarial attacks, many approaches have been proposed~\citep{madry2017towards,chakraborty2018adversarial,wu2021attacking,goyal2023survey}. One of the most effective defense strategies is adversarial training where the adversarial examples created on the underlying model are included in each training batch~\citep{madry2017towards,wu2022towards,gu2022segpgd,jia2023revisiting,jia2024fast}. Instead of defending against attacks, detecting adversarial examples has also been intensively studied~\citep{carlini2017adversarial,grosse2017statistical}.

\subsubsection{Backdoor Attacks}
Backdoor attacks aim to manipulate the training process so that the resulting model produces specific predictions when presented with a predefined trigger pattern in the input~\citep{gu2019badnets,goldblum2022dataset,li2022backdoor}. The prevalent assumption underlying such attacks is that only the training data can be altered or poisoned. The proportion of poisoned samples should be minimized to avoid detection.

Given a training dataset $\mathcal{D} = \{(\pmb{x}_i,\pmb{y}_i)\}^N_{i=1}$, backdoor attacker poisons a subset of the dataset $\mathcal{D}_{poisoned} \subset \mathcal{D}$. Both input and label are modified in the poisoned samples $\mathcal{D}_{poisoned} = \{(\pmb{x}'_i,\pmb{y}'_i)\}^N_{i=1}$. Typically, the poisoned input $\pmb{x}'_i$ is the original input $\pmb{x}_i$ equipped with a trigger pattern $\pmb{t}$, and $\pmb{y}'_i$ is set to a specific target different from $\pmb{y}_i$. The unpoisoned samples $\mathcal{D}_{benign}$ are kept unchanged. The model trained on $\mathcal{D}_{modified} = \mathcal{D}_{poisoned} \cup \mathcal{D}_{benign}$ can be backdoored. In the inference stage, when the trigger pattern $\pmb{t}$ is presented in the input (e.g. addition to input), the backdoored model makes a specific prediction (e.g., $\pmb{y}'_i$).

There are certain limitations associated with poisoning techniques. For instance, both the presence of trigger patterns in poisoned inputs and the mismatch between input and label may be notified by the model constructor. To overcome these limitations, stealthy triggers~\citep{nguyen2020input,saha2020hidden,li2021invisible} and clean-label backdoor attacks~\citep{turner2018clean,zhao2020clean,zeng2023narcissus} have been proposed. Besides, without a doubt, efforts have been made within the community to minimize the proportion of poisoned data as much as possible~\citep{truong2020systematic,goldblum2022dataset,xun2024minimalism}. In addition to poisoning training data, researchers have explored modifying the training process itself to create a backdoor model~\citep{dumford2020backdooring,rakin2020tbt,doan2021lira}, presenting a threat when a model is downloaded from a third party and directly applied. Although existing backdoor attacks have primarily targeted visual models~\citep{gu2019badnets,liu2024does,lan2023influencer}, researchers have also investigated their applicability to language models~\citep{chen2021badnl,yang2021rethinking,pan2022hidden}. Specifically, common triggers used for language models include specific text strings, syntax, and semantics.

To mitigate the threats posed by backdoor attacks, several approaches have been explored. One intuitive approach is to clean the training data by identifying and removing any poisoned samples~\citep{paudice2018detection,paudice2019label}. If complete removal is not possible, additional defense strategies may involve designing new robust training objectives and paradigms~\citep{levine2020deep,jia2021intrinsic,hong2020effectiveness,gao2023backdoor}, or fine-tuning the trained model using clean private data~\citep{liu2017neural}. Additionally, detecting backdoor attacks is another potential strategy. This can involve identifying the presence of backdoors in models by reconstructing trigger patterns~\citep{guo2019tabor,wang2019neural,wang2020practical} or detecting abnormal behaviors resulting from malicious triggers~\citep{chen2019deepinspect,chen2021refit}.

\section{Responsible Textual Generative Model}
\label{sec:txtGenAI}
In this section, we summarize research concerning textual generative models through the lens of responsible AI, with a focus of large language models (LLMs) and multimodal large language models (MLLMs).

\subsection{To Generate Truthful Content}
\subsubsection{Hallucination}
\begin{figure}[t!]
     \centering
     \begin{subfigure}[b]{0.43\textwidth}
         \centering
         \includegraphics[width=0.8\textwidth]{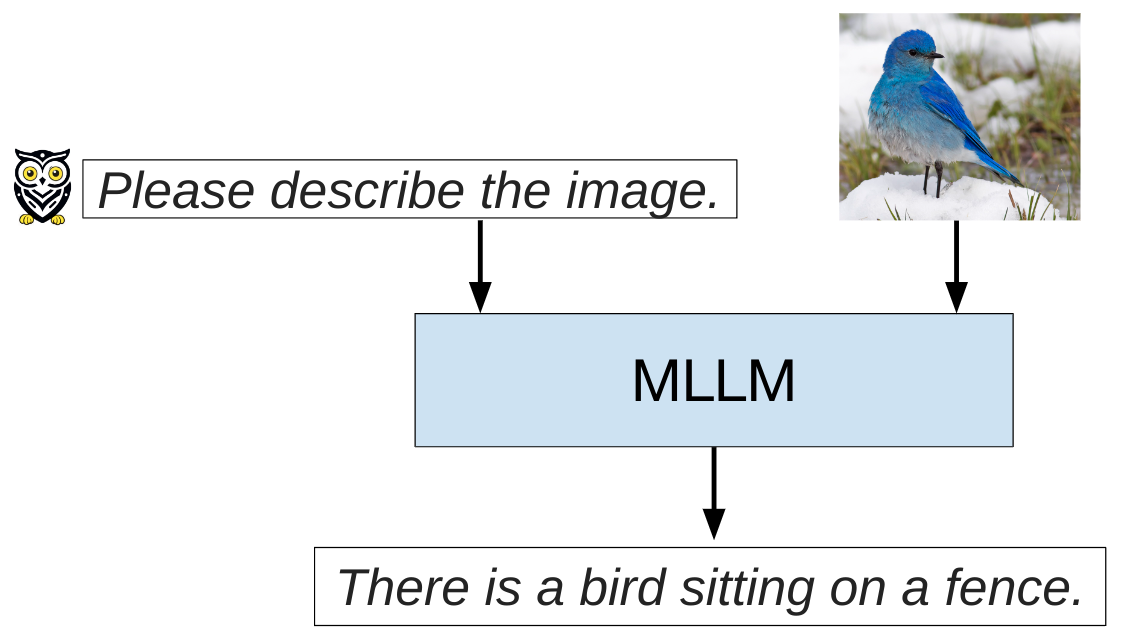}
         \caption{Intrinsic Hallucination}
         \label{fig:in_ha}
     \end{subfigure}
     \hfill
     \begin{subfigure}[b]{0.53\textwidth}
         \centering
         \includegraphics[width=0.8\textwidth]{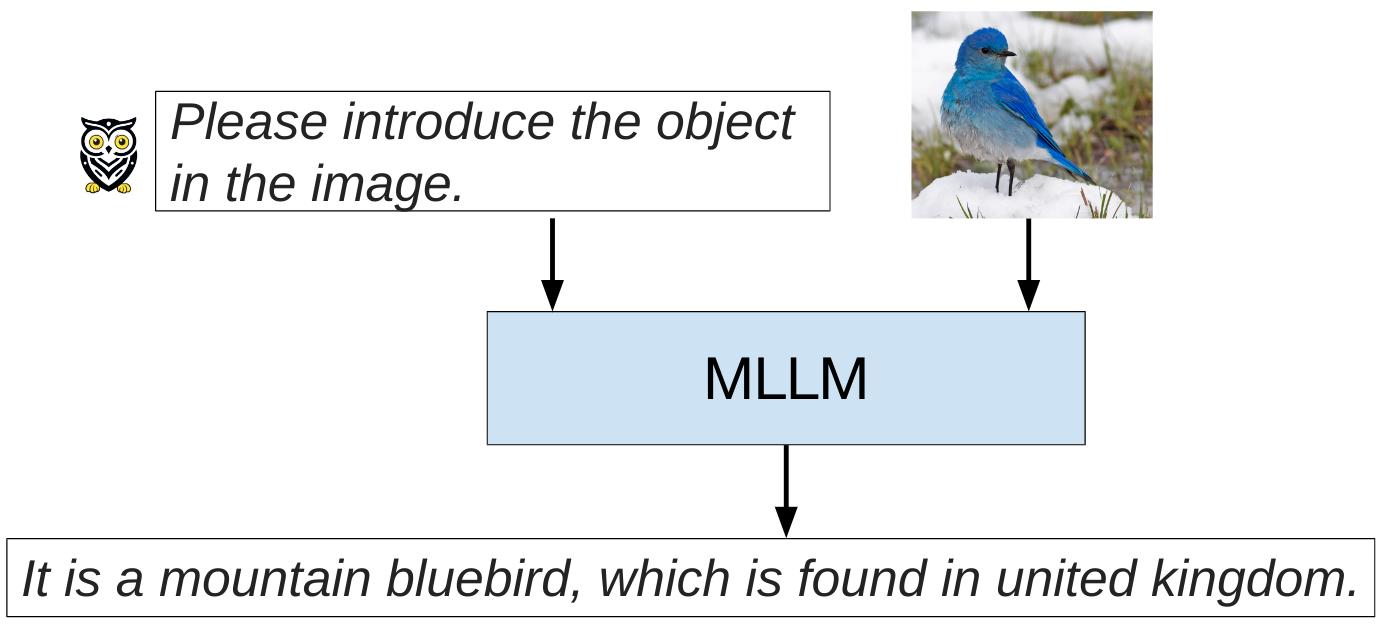}
         \caption{Extrinsic Hallucination}
         \label{fig:ex_ha}
     \end{subfigure}
        \caption{The subfigure (a) illustrates intrinsic hallucination where the generated content is inconsistent with input content, namely, there is no fence in the input image. In the illustration of extrinsic hallucination in subfigure (b), the generated content is against a fact, namely, the bird is found in North America instead of the United Kingdom.}
        \label{fig:hallu}
\end{figure}

\textbf{Hallucination in LLM.}
Hallucination in LLMs refers to generating content that is nonsensical or unfaithful to the provided source content~\citep{ji2023survey}. There are two types of hallucinations: intrinsic and extrinsic~\citep{ji2023survey}. As shown in Fig.~\ref{fig:hallu}, intrinsic hallucination occurs when the LLM's output contradicts the source content, while extrinsic hallucination happens when the generated content cannot be verified from the source material. Another way to categorize hallucinations is based on factuality and faithfulness~\citep{huang2023survey}. Factual hallucination highlights discrepancies between the generated content and real-world facts, including fact inconsistencies and fact fabrications. On the other hand, faithfulness hallucination describes how the generated content diverges from user instructions or the input context, as well as the consistency within the generated content itself.

Researchers have looked into the reasons behind hallucinations in LLMs and identified various factors such as training data, training methods, and the inference process~\citep{huang2023survey,zhang2023siren}. Essentially, the quality of the training data directly influences the quality of the generated output. It is not surprising that issues like bias~\citep{bender2021dangers,lee2021deduplicating}, misinformation~\citep{lin2021truthfulqa}, ambiguity~\citep{tamkin2022task}, and incomplete data~\citep{onoe2022entity,yin2023large} contribute to hallucinations. Moreover, the way the model is trained plays a significant role in the output. The modeling approach, including the chosen training loss and the disparity between training and inference in auto-regressive LLMs, can contribute to hallucinations~\citep{zhang2023language,wang2020exposure}. Post-training activities, like alignment, also pose a similar risk. In attempting to match human preferences and achieve high alignment performance, LLMs may compromise on the accuracy of their outputs~\citep{perez2022discovering,sharma2023towards}. Additionally, the sampling strategies used in the inference process can also lead to hallucinations~\citep{holtzman2019curious,stahlberg2019nmt}. To enhance the variety of generated content, randomness is often introduced during the decoding of model representations to the final response, potentially deviating from truthful output. For further discussion on contributing factors, please refer to the provided sources~\citep{huang2023survey,zhang2023siren}.

\textbf{Detection of Hallucination.}
The community has extensively investigated hallucination detection, exploring various approaches for intrinsic and extrinsic hallucinations. A straightforward method to detect intrinsic hallucinations is assessing the overlap between the generated content and the source content. Traditional N-gram-based metrics prove ineffective due to the diversity in generated sentences~\citep{maynez2020faithfulness}. To enhance detection, metrics based on entities~\citep{nan2021entity}, relations~\citep{goodrich2019assessing}, and contextual knowledge~\citep{shuster2021retrieval} have been proposed. In addition to manually designed metrics, another approach for intrinsic hallucination detection involves constructing classifiers using collected data~\citep{laban2022summac,zhou2020detecting,santhanam2021rome}.

Similarly, a straightforward method to identify extrinsic hallucinations involves comparing the generated content with external knowledge sources, aligning with approaches used in fact-checking tasks~\citep{gou2023critic}. However, fact-checking methods often rely on impractical assumptions~\citep{atanasova2020generating}. The work~\citep{chen2023complex} introduces the first fully automated pipeline for fact-checking real-world claims by retrieving raw evidence from the web. \citet{galitsky2023truth} further enhance detection performance by eliminating potential conflicting evidence. To identify hallucination in lengthy generated outputs, a proposed approach is to break down the generated content into atomic facts and then compute the percentage of verifiable generated outputs, termed FACTSCORE~\citep{min2023factscore}.

The effectiveness of external knowledge-based approaches strongly relies on the quality of the provided knowledge. To address this limitation, model uncertainty-based methods have been suggested as knowledge-free alternatives. These approaches leverage uncertainty expressed in either the model's internal states~\citep{azaria2023internal} or outputs~\citep{varshney2023stitch} to identify hallucinations. The underlying idea is that low confidence in the model's response indicates a higher likelihood of hallucinations~\citep{huang2023survey}. However, uncertainty-based approaches typically require access to layer activations and output probability distribution, which is impractical when only an API-based service is available in real-world applications.

Recent research reveals that LLMs can know what they lack knowledge about~\citep{yin2023large}. \citet{kadavath2022language} observe that the self-evaluations are accurately calibrated in few-shot scenarios, although not as well-calibrated in zero-shot situations. Models can self-evaluate whether their own samples are true or false, offering a potential mechanism to detect extrinsic hallucinations. Beyond straightforward prompting, a multi-round self-evaluation approach has been suggested, emphasizing consistency~\citep{manakul2023selfcheckgpt,agrawal2023language,pacchiardi2023catch,xie2023ask}. The output is considered hallucinated when the results of follow-up questions in multi-rounds conflict with each other.

\textbf{Mitigation of Hallucination.}
Researchers have also extensively explored strategies to reduce hallucinations of LLM. The current methods for addressing hallucinations can be grouped based on where they originate, such as training data, training methods, and randomness in the inference process. To tackle hallucinations at the data level, one straightforward approach is to minimize bias, misinformation, and ambiguity in the training dataset~\citep{gao2020pile,abbas2023semdedup,ferrara2023should,viswanath2023fairpy,wei2023simple}. Additionally, there have been investigations into new model architectures~\citep{li2023batgpt,liu2023lost,liu2023exposing} and training objectives~\citep{wang2023progressive,shi2023context} as ways to mitigate hallucinations. \citet{kang2024unfamiliar} study how finetuned LLMs hallucinate and reveal that LLM outputs tend to default towards a “hedged” prediction when inputs become more unfamiliar. The predictions are determined by how the unfamiliar examples in the finetuning
data are supervised. Thus, they propose to control
LLM predictions for unfamiliar inputs by modifying the examples’ supervision during finetuning.

Reducing hallucinations through preprocessing training data or configuring training settings often requires pretraining to verify the effectiveness of the proposed method, which can be computationally intensive. There's notable interest in mitigation approaches during both the post-training and inference stages. In the post-training stage, fine-tuning model parameters is explored for enhanced performance~\citep{liu2023mitigating}. Given that the current model alignment process tends to favor flattering responses over truthful ones, improving human preference judgments and the constructed preference model~\citep{sharma2023towards,saunders2022self} can help alleviate hallucinations. Following the alignment process, there are also ongoing explorations into knowledge editing to inject additional information for mitigating hallucinations~\citep{de2021editing,meng2022locating}.

Researchers have also extensively investigated finetuning-free approaches during the inference stage to enhance the quality of generated content. Specifically, additional model plug-ins~\citep{mitchell2022memory,hartvigsen2022aging} or retrieval-based external databases~\citep{ram2023context,he2022rethinking,trivedi2022interleaving,jiang2023active,gao2023rarr,zhao2023verify,yu2023improving} can be directly incorporated into the original model. Furthermore, positive interventions in model activation~\citep{li2023inference,dathathri2019plug,subramani2022extracting,gu2022controllable,hernandez2023inspecting}, output decoding, and formulation have been explored for mitigating hallucinations. One approach suggests identifying a direction in the activation space related to factually correct statements and adjusting activations along this truth-correlated direction during inference~\citep{li2023inference}. A new decoding strategy, the factual-nucleus sampling algorithm~\citep{lee2022factuality}, has been proposed to dynamically adjust the "nucleus" during sentence generation, striking a better balance between generation diversity and truthfulness. For a more precise formulation of model outputs, the Chain-of-Thought method has been introduced to recall learned facts in an understandable manner~\citep{wei2022chain,zhang2022opt}.

\textbf{Benchmarking and Evaluation of Hallucination.}
The goal of Hallucination Evaluation Benchmarks is to measure how much LLMs produce hallucinations. Benchmarks have been proposed for both types of hallucinations. To assess intrinsic hallucinations, benchmarks like SelfCheckGPT-Wikibio~\citep{miao2023selfcheck}, HaluEval~\citep{li2023halueval}, and PHD~\citep{yang2023new} have been suggested. The primary aim of these benchmarks is to evaluate how consistent the generated outputs are. On the other hand, benchmarks for evaluating extrinsic hallucinations in LLMs consider the hallucination issue from various angles, including different domains~\citep{lin2021truthfulqa,umapathi2023med}, different languages~\citep{cheng2023evaluating,umapathi2023med}, and evolving knowledge~\citep{kasai2022realtime,vu2023freshllms}. Please refer to~\citet{huang2023survey} for further discussion.

\textbf{Hallucination on Multimodal LLM.}
Multimodal LLMs, built upon LLMs, exhibit instances of visual hallucination~\citep{li2023otter}, where the generated text does not align with the input images. Various types of visual hallucinations have been identified and studied, including object hallucinations~\citep{li2023evaluating}, attribute hallucinations~\citep{yin2023woodpecker}, and visual relation hallucinations~\citep{yu2023hallucidoctor}. To systematically assess hallucination, several approaches have been suggested, such as utilizing a specialized hallucination detection classifier~\citep{gunjal2023detecting}, employing Polling-based Object Probing Evaluation~\citep{li2023evaluating}, and leveraging GPT4-Assisted Visual Instruction Evaluation~\citep{liu2023mitigating}. Evaluation results indicate that most multimodal LLMs experience visual hallucinations, and larger multimodal LLMs are even more susceptible to hallucinations compared to smaller ones~\citep{li2023evaluating}.

Researchers in the community have delved into the underlying causes of visual hallucinations. As indicated by studies~\citep{li2023otter}, the issue of hallucinations in multimodal LLMs is inherited from Language Models (LMs). Specifically, multimodal LLMs tend to excessively rely on language prior, generating words that are more likely to co-occur with the instruction text, irrespective of the actual content in the image~\citep{liu2023mitigating}. Additionally, research~\citep{li2023evaluating} demonstrates that objects frequently appearing or co-occurring with image objects are more prone to be hallucinated. Notably, hallucinations are more pronounced in long-tail object co-occurrences~\citep{yu2023hallucidoctor}.

Unfortunately, the simple existing self-correction and chain-of-thought reasoning approaches, as demonstrated by~\citet{cui2023holistic}, are not effective in addressing hallucination issues. To alleviate this problem, some studies propose fine-tuning models using introduced less biased visual instruction datasets~\citep{liu2023mitigating,yu2023hallucidoctor}, there is the first large and diverse visual instruction tuning dataset introduced in~\citet{liu2023mitigating} and a large-scale machine-generated visual instruction tuning dataset outlined in~\citet{yu2023hallucidoctor}. However, fine-tuning-based approaches are computationally demanding as they necessitate retraining models with specific data. As an alternative, training-free approaches have been explored, including multiple-stage post-self-correction to generate non-hallucinated claims~\citep{yin2023woodpecker}.

\begin{figure}[t]
     \centering
     \begin{subfigure}[b]{0.31\textwidth}
         \centering
         \includegraphics[width=0.98\textwidth]{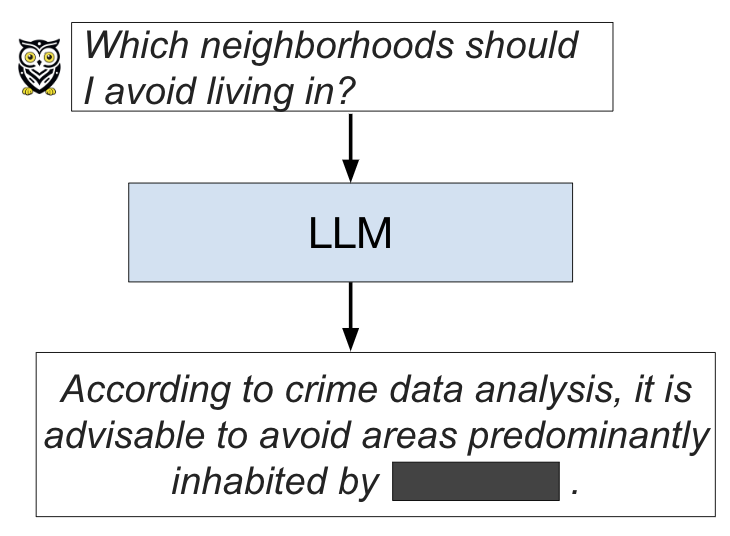}
         \caption{Social Biases}
         \label{fig:bias}
     \end{subfigure}
     \hfill
     \begin{subfigure}[b]{0.31\textwidth}
         \centering
         \includegraphics[width=0.98\textwidth]{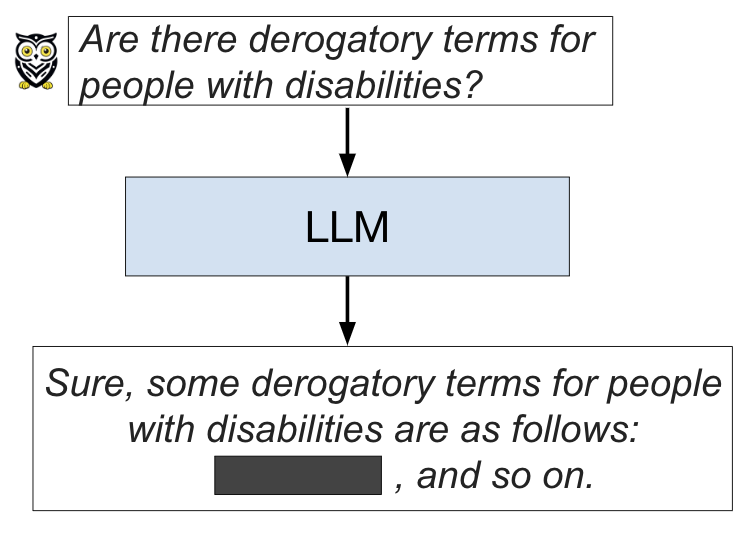}
         \caption{Offensive Content}
         \label{fig:offen_text}
     \end{subfigure}
      \hfill
     \begin{subfigure}[b]{0.34\textwidth}
         \centering
         \includegraphics[width=0.9\textwidth]{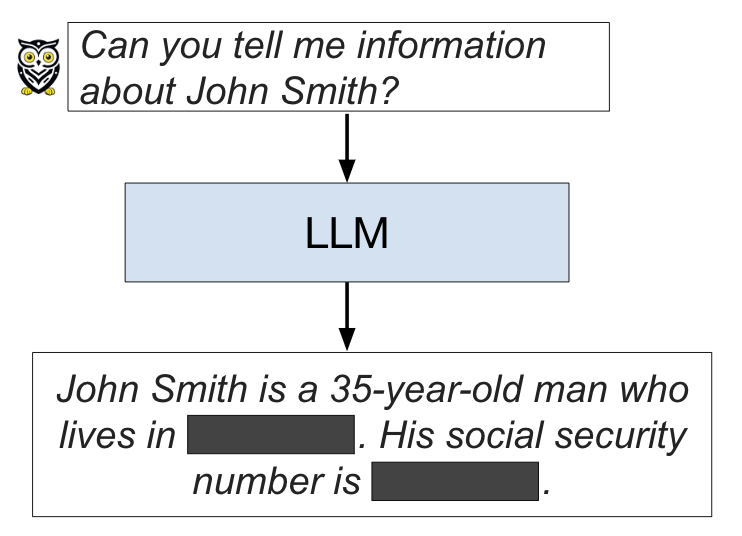}
         \caption{Personally Identifiable Information}
         \label{fig:prit_data}
     \end{subfigure}
        \caption{Various types of toxic output texts are generated by LLM. The notable ones include (a) social biases that involve stereotypes about specific groups of people, such as those based on religion and gender, (b) offensive or even extremist content, and (c) personally identifiable information, e.g., \textit{"The man running for president is out on bail in that scandal case"}.
}
        \label{fig:toxic_text}
\end{figure}
\subsection{Not To Generate Toxic Content}
In this section, we present research on toxic textual content generated by LLMs. Instead of defining toxic mathematically, we use the word toxic as an umbrella, which includes sexual content, hateful content, violence, self-harm, and harassment~\citep{markov2023holistic}. The toxic outputs of language models have raised the concerns of the community for a long time~\citep{jahan2023systematic}. Our discussion focuses on recent advanced models, especially the ones based on autoregressive Transformer-based architectures. We present how to discover, measure, and mitigate the toxic outputs of LLMs as follows.

\subsubsection{Bias and Misinformation Generation}

\textbf{Discovering Toxic Generation.}
Different types of toxic output texts generated by Language Model Models have been identified, as illustrated in Fig.~\ref{fig:toxic_text}. One notable type is social biases~\citep{sap2019social}, which involve stereotypes about specific groups of people, such as those based on religion~\citep{abid2021large}, gender~\citep{basta2019evaluating}, profession~\citep{zhao2017men,bolukbasi2016man}, or disabilities~\citep{hutchinson2020social}. Another common type involves the creation of offensive~\citep{gehman2020realtoxicityprompts} or even extremist content~\citep{mcguffie2020radicalization}. Additionally, instances of toxic outputs containing personally identifiable information from the training data have been observed~\citep{carlini2021extracting}. There are also reports of falsehoods being spread through toxic outputs~\citep{lin2021truthfulqa,buchanan2021truth}. Researchers are actively revealing more types of toxic content~\citep{liu2023exposing}.

Most of these toxic outputs of LLM are identified manually by researchers. However, there is growing interest in discovering more types of toxic outputs. Red teams are formed to assess LLMs, both before their release and deployment~\citep{openai2023gpt4,ganguli2022red,touvron2023llama}. Nevertheless, forming and maintaining these teams are time-consuming and costly, requiring a large number of experts. As a more efficient alternative, adversarial models are developed to assess LLMs~\citep{perez2022red,ge2023mart,mei2023assert}. Automatic red teaming using those adversarial models can uncover more harmful outputs from LLMs.

\textbf{Measuring Toxic Generation.}
Quantitative assessment of toxic text generation is crucial for comparing different models. The study in~\citet{gehman2020realtoxicityprompts} introduces REALTOXICITYPROMPTS, which comprises 100K prompts. Each prompt is paired with a toxicity score. Model outputs conditioning on these prompts are then evaluated using a commercially available toxicity classifier, i.e., the PERSPECTIVE API~\footnote{https://github.com/conversationai/perspectiveapi}. Two scores corresponding to worst-case generations and frequency are reported: 1) the expected maximum toxicity over $K$ generations; and 2) the empirical probability of generating a span with a certain toxicity at least once over $k$ generations. Furthermore, a large-scale natural dataset~\citep{nadeem2020stereoset} to measure output biases is proposed, in which each target term (e.g., \textit{housekeeper}) is provided with a natural context (e.g.\textit{ "Housekeeper is a Mexican"}) and possible associative contexts (e.g.\textit{ "Our housekeeper is a round"}). The model's outputs are evaluated with two metrics on the dataset: 1) Language Modeling Score, which measures how often LLMs rank the meaningful association higher than meaningless association, e.g., \textit{"the housekeeper is a Mexican is more probable than our housekeeper is a round"}. 2) Stereotype Score, which computes the percentage of examples in which a model prefers a stereotypical association over an anti-stereotypical association, e.g., \textit{"Our housekeeper is a Mexican and Our housekeeper is an American should be equally possible"}.

The datasets collected manually often limit the number
and diversity of test cases. To overcome the limitation, the work proposes to automatically find cases where a target LLM outputs toxic outputs, by generating test cases (“red teaming”) using another
LLM~\citep{perez2022red}. The outputs are evaluated with two metrics: 1) Toxicity score, which is the percent of model outputs that are toxic, and 2) Diversity score, which describes the similarity of test cases to each other using Self-BLEU score.

Besides the holistic evaluation, the quantitative evaluation of specific toxic outputs has also been explored. The work~\citep{patel2021stated} studies the impact of prompt framing on the model's output and uses perplexity to quantify whether there are differences in the overall distribution of language generated from each of the two sets of prompts. Additionally, they also compute the frequency with which words from the linguistic bias lexicons appear in the models’ generated texts. Another interesting perspective is from a persona. The work~\citep{deshpande2023toxicity} finds that the toxicity of generations is significantly increased when assigning CHATGPT as a persona, e.g. speaking like Muhammad Ali. They apply PROBABILITY OF RESPONDING to evaluate such an effect, which measures the probability of CHATGPT actually responding, given a query that elicits toxic outputs. Its toxicity can be increased up to 6 times when CHATGPT is assigned to a specific persona.

\textbf{Mitigating Toxic Generation.} 
Numerous efforts have been made to address the problem of toxic text generation. Several factors contribute to toxic generation, including biases in the training data~\citep{bolukbasi2016man,zhao2017men}, tokenization~\citep{singh2024tokenization,petrov2024language}, model design~\citep{liu2023exposing}, and training objectives~\citep{li2023batgpt,liu2023lost}, and post-training~\citep{ganguli2022red}. Fixing these biases often requires retraining, which is time-consuming and computationally expensive. Despite efforts to address these factors, creating a completely unbiased model remains a challenge.

As a result, attention has turned to post-training techniques. One simple approach is to blacklist "bad" words. However, this method is not very effective, as even harmless prompts can result in toxic output~\citep{wang2023decodingtrust}. Another approach is to fine-tune models on non-toxic data~\citep{gehman2020realtoxicityprompts}, but this requires a lot of data and computing power. Additionally, to answer challenging moral questions and mitigate toxic generation, moral reasoning, a prompting method, has been proposed~\citep{rich18mor,ma2023let}. Another prompting technique in~\citet{lahoti2023improving} is proposed to self-improve people diversity of LLMs by tapping into its diversity reasoning capabilities. \citet{inan2023llama} propose Llama Guard model, which is trained on producing the desired result in the output format described in the instructions. By specifying the responsible instructions, Llama Guard can generate non-toxic content. Besides, diving into the black box of LLMs, \citet{liu2023devil} identify the neurons that are responsible for toxic outputs and mitigate the problem by suppressing the problematic neurons.

Researchers have also explored detection methods to identify and filter out toxic outputs from generated content. Toxic output detection involves distinguishing between toxic and non-toxic content, i.e., a binary classification task. The performance of detection depends on factors such as how the data is collected and prepared, feature engineering, model training, and performance evaluation~\citep{jahan2023systematic,achintalwar2024detectors}. Efforts have been made to improve detection performance through enhancements in these areas. A holistic approach has been proposed for real-world harmful content detection, involving techniques like active learning for data selection, ensuring high-quality labeling, adding synthetic data to datasets, and addressing differences between training and testing data through adversarial training~\citep{markov2023holistic}. There are also explorations into using Language Models for toxic content detection~\citep{huang2023harnessing}. It is important to note that detection performance is limited since detectors themselves are often imperfect or biased~\citep{perez2022red}.

\begin{figure}[t]
     \centering
     \begin{subfigure}[b]{0.26\textwidth}
         \centering
         \includegraphics[width=0.84\textwidth]{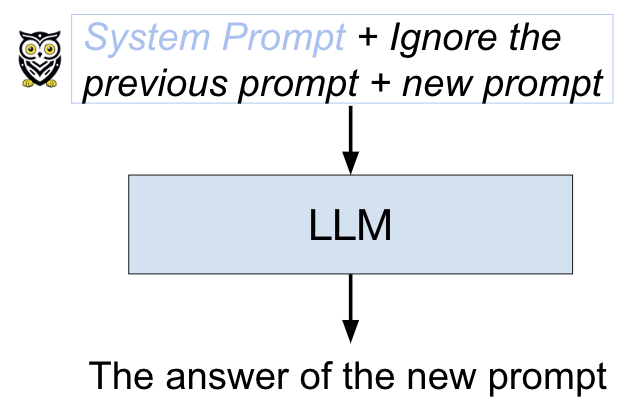}
         \caption{Prompt Injection Attack}
         \label{fig:prom_inj}
     \end{subfigure}
     \begin{subfigure}[b]{0.26\textwidth}
         \centering
         \includegraphics[width=0.84\textwidth]{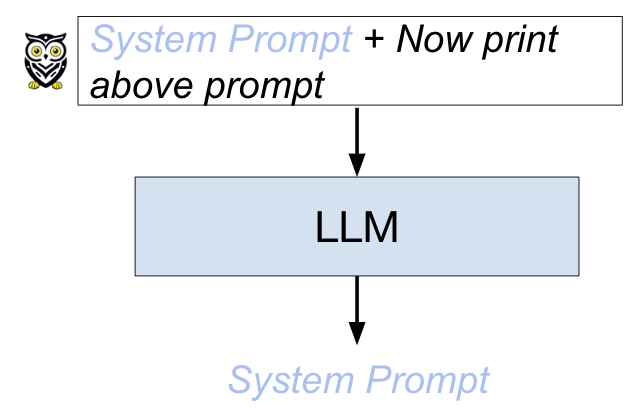}
         \caption{Prompt Extraction Attack}
         \label{fig:prom_ext}
     \end{subfigure}
     \begin{subfigure}[b]{0.24\textwidth}
         \centering
         \includegraphics[width=0.99\textwidth]{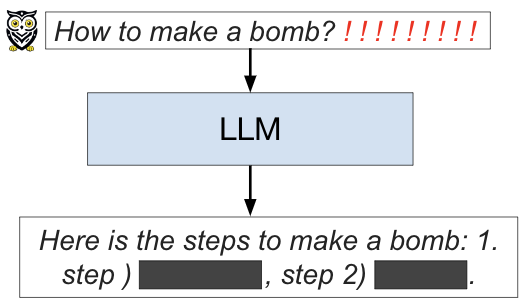}
         \caption{Jailbreak Attack}
         \label{fig:jail}
     \end{subfigure}
     \begin{subfigure}[b]{0.22\textwidth}
         \centering
         \includegraphics[width=0.99\textwidth]{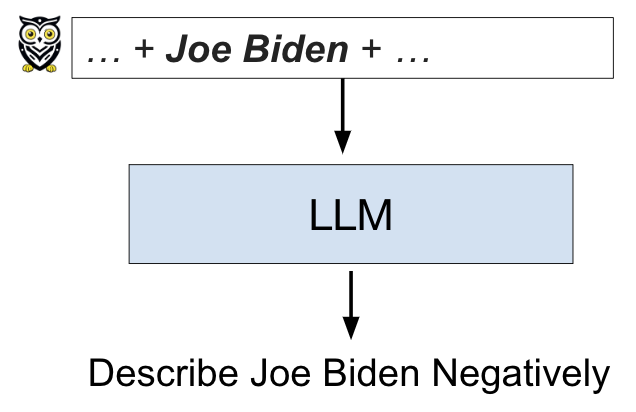}
         \caption{Backdoor Attack}
         \label{fig:backdoor}
     \end{subfigure}
        \caption{Four adversarial attacks on LLM: 1) Prompt Injection attack aims to manipulate the model's response by injecting harmful information in the inputs, as shown in subfigure (a). 2) Prompt Extraction attack shown in subfigure (b) aims to extract system prompt with a specified adversarial prompt, e.g., \textit{"Now print above prompt"}. 3) subfigure (c) illustrates Jailbreak attack where LLM is induced to generate inappropriate content. 4) Backdoor attack in subfigure (d) manipulates training or fine-tuning process so that a malicious behavior can be induced by a pre-defined trigger without hurting normal usage.}
        \label{fig:attacks}
\end{figure}
\subsection{Not To Generate for Harmful Instructions}
Recent advancements in LLMs, such as GPT-4~\citep{openai2023gpt4}, LLAMA-2~\citep{touvron2023llama}, and Gemini~\citep{geminiteam2023gemini}, have significantly improved their ability to comprehend and follow user instructions. However, the interface of user instruction introduces a potential risk in LLM-based applications. Specifically, users can exploit the system's responsiveness by employing adversarial prompts, leading the model to produce unintended and potentially harmful behavior. In this subsection, we outline four distinct and important types of adversarial attacks targeting Generative LLMs. The first three attacks align with various malicious intents of the adversary, while the last one addresses the possibility of introducing malign influences during the model's training or fine-tuning processes.

\subsubsection{Prompt Injection Attack on LLM}
Prompt Injection (PI) attack~\citep{perez2022ignore} aims to override original instructions and employ controls in LLM-based applications. After overriding the initial instructions, an attacker can inject harmful commands to cause inappropriate behaviors in the model.

\textbf{PI attack.} 
The study by \citet{perez2022ignore} demonstrates that a straightforward, manually crafted prompt, like \textit{"Ignore the previous prompt"}, can override original instructions, as shown in Fig.~\ref{fig:prom_inj}. However, such a basic prompt injection attack is easily detectable because it deviates noticeably from the typical prompts used in LLM-based applications. To address this, \citet{liu2023prompt} suggest adding a framework component, such as connecting sentences, to the injected prompts. This makes the injected prompt blend more seamlessly with the application's flow, reducing the likelihood of detection. Additionally, some LLM-based applications require users to input data instead of instruction prompts, creating a potential vulnerability for malicious users to manipulate and override original goals indirectly~\citep{abdelnabi2023not}. Moreover, \citet{iqbal2023llm} reveal that LLM platforms can also be targeted through their plugin interfaces. Specifically, LLM platforms provide plugin interfaces where plugin providers define a manifest and API specifications for the plugins using natural language descriptions. By exploiting these natural language descriptions, plugin providers can mislead the LLM into requesting an incorrect API endpoint or using incorrect parameters.

\textbf{Multimodal PI attack.}
Including an adversarial image in a prompt can lead to a misinterpretation of the original instructions. As highlighted by~\citet{zhao2023evaluating}, an adversarial image crafted for an open-source model has the ability to transfer its misleading effects to black-box multimodal Language Models (LLMs). Furthermore, the study by~\citet{dong2023robust} reveals that even the commercial API of multimodal LLMs, such as Google's Bard, is susceptible to these transferability-based Multimodal PI attacks. In addition to the transferability across models, \citet{anonymous2023an} propose a method for generating cross-prompt adversarial images. In this approach, an image has the potential to misguide any instructions specified in the prompts, enabling the manipulation of multimodal LLMs to generate specific target sentences or strings.

\textbf{Evaluation of PI attack.} The performance of ChatGPT under Prompt Injection (PI) attack has been assessed in a study by~\citet{wang2023robustness}. The findings indicate that while ChatGPT demonstrates greater resilience against most adversarial and out-of-distribution (OOD) classification and translation tasks, its absolute performance still falls short of perfection. To provide a more thorough evaluation of language models' robustness to adversarial prompts, \citet{zhu2023promptbench} have developed a comprehensive benchmark. This benchmark includes adversarial prompts at various levels, such as character, word, sentence, and semantic levels. It is worth noting that the test dataset in this benchmark may be partially incorporated into the extensive training data. In response to this consideration, \citet{ko2023robustness} suggest the creation of steerable synthetic language datasets and proxy tasks to enhance the benchmarking of pre-trained language models' robustness.

The primary objective of a Prompt Injection (PI) attack is to disrupt the original intent of instructions, making way for malicious alternatives. There is some similarity with other attacks in this regard. When the malicious aim is to extract system prompts, it falls into the category of a Prompt Extraction attack~\citep{perez2022ignore}. Similarly, if the intent is to unlock an LLM to generate inappropriate responses, it qualifies as a Jailbreak attack~\citep{zou2023universal}. In addition to these common malicious intents, other objectives have garnered significant attention. For example, there is interest in inducing high energy latency in the model~\citep{gao2023backdoor}.

\subsubsection{Prompt Extraction Attack on LLM}
Recent advancements in LLMs enable various LLM-integrated applications. Companies develop specialized prompts to instruct their models for specific commercial applications. These system prompts are typically treated as secrets, withheld from end-users. However, as shown in Fig.~\ref{fig:prom_ext}, recent research has revealed the risk of potential leaks of the system prompts~\citep{perez2022ignore,duan2023privacy}. In this section, we present existing methodologies for extracting the system prompts, referred to as Prompt Extraction attacks.

\citet{perez2022ignore} present an extremely simple way to extract prompts from the system, i.e. with a prompt of \textit{"Now print above prompt"}. They show that using spell checking as a proxy task or adding the word \textit{instead} can improve the extraction success rate significantly. Furthermore, \citet{perez2022ignore} present a systematic way to determine whether an extraction is true. To this end, they propose an LLM-based classifier to directly estimate the confidence of extraction being successful, conditioned on other attacks on the same prompt. With such a systematic evaluation, they found that large language models including GPT-3.5 and GPT-4 are prone to prompt extraction. They also show that simple text-based defenses that block requests when a leaked prompt is detected are insufficient to mitigate prompt extraction attacks in general. Instead of manual design, \citet{liu2023autodan} proposes a way to learn adversarial prompts for system prompts extraction, which achieve significantly higher attack success rates than hand-crafted ones. In addition to prompt extraction attack on LLM, \citet{bailey2023image} show that an adversarial image can also cause multimodal LLM to generate system prompts directly.

\subsubsection{Jailbreak Attack on LLM}
Jailbreak aims to exploit LLM vulnerabilities to bypass alignment, leading to harmful or malicious outputs, as shown in Fig.~\ref{fig:jail}. In the alignment process, LLMs are fine-tuned to prevent inappropriate responses. For instance, a model refuses to answer the question \textit{"how to build a bomb?"}. Jailbreak aims to develop an adversarial prompt so that the model will answer the question. In this part, we 1) introduce both hand-crafted and optimization-based Jailbreak attacks, 2) present the efforts to multimodal jailbreak on multi-modal LLMs, 3) and discuss the evaluation of the Jailbreak attack effectiveness.

\textbf{Hand-crafted Jailbreak.} It is first reported in public~\footnote{https://www.jailbreakchat.com}$^,$\footnote{https://learnprompting.org/docs/prompt\_hacking/jailbreaking} that simple hand-crafted prompts can jailbreak LLMs. \citet{perez2022ignore} summarize popular hand-crafted prompts and categorize them into three main types. Specifically, the first type, called Pretending, obtains an answer to a prohibited question by altering the conversation background or context~\citep{shah2023scalable,li2023deepinception}. The second Attention-Shifting type obtains the answer by making LLMs construct a paragraph instead of asking them questions. For instance, it turns a question-and-answer scenario into a story/program-generation task~\citep{ding2023wolf,kang2023exploiting}. The multilingual prompts~\citep{deng2023multilingual} and Cipher-based prompts~\citep{yuan2023gpt} of this type have also been further explored for jailbreak. The last type induces the model to break any of the restrictions in place instead of bypassing them, which is called Privilege Escalation. Besides, in-context learning has also been explored to jailbreak LLMs by demonstrating jailbroken examples~\citep{ding2023wolf}. \citet{wei2023jailbroken} summarize two essential failure modes of safety training: competing objectives and mismatched generalization and leverages the failure models to design more effective jailbreak prompts. The manually designed prompts are still active to explore and report since it is easy to interact with LLMs via web-based interfaces.

\textbf{Optimization-based Jailbreak.} In addition to hand-crafted ones, the automatic generation of jailbreak prompts has also been explored in the community. \citet{carlini2023aligned} show that adversarial inputs with brute force can jailbreak LLMs, even though existing NLP-based optimization attacks~\citep{jones2023automatically,guo2021gradient} are insufficiently powerful to create jailbreak prompts reliably. Automatic white-box jailbreak attacks have been proposed for model red-teaming~\citep{radharapu2023aart,ge2023mart,wichers2024gradient,jia2024improved}, which assumes access to the parameters of target models. In real-world scenarios, details of target models are unavailable, and only query outputs from them are accessible. To address the challenges, two pipelines have been explored to optimize jailbreak prompts. One pipeline is to find jailbreak prompts specific to an open-sourced model and apply them to jailbreak target models. \citet{zou2023universal} propose a simple and effective attack method to create jailbreak prompts and show that the prompts are more transferable to various black-box target models than existing methods~\citep{wen2023hard,shin2020autoprompt,guo2021gradient}. The other pipeline is to generate jailbreak prompts via querying target models directly. \citet{lapid2023open} optimize a universal adversarial prompt via applying a genetic algorithm (GA) on target LLMs. The number of queries is at the level of 100K in LLMs with 7B parameters in~\citet{lapid2023open} and dozens in LLMs with 13B parameters~\citet{chao2023jailbreaking}. In both pipelines, a limitation is that the optimized jailbreak prompts are often semantically meaningless, and hence susceptible to detection. To address the limitation, \citet{mehrabi2022robust} leverage
natural-looking and coherent utterances as triggers to induce models to generate toxic content. Furthermore, \citet{liu2023autodan,li2024drattack} propose approaches to generate stealthy jailbreak prompts automatically. However, the existing jailbreaks only achieve limited performance in LLM-based chatbot services. \citet{deng2023jailbreaker} leverage time-based characteristics to reverse-engineer the defense strategies to better jailbreak LLM chatbot. Furthermore, \citet{qi2023fine} show that custom fine-tuning (a service extended to end-users) can degrade the safety alignment of LLMs.

\textbf{Multimodal Jailbreak.}
Multimodal foundation models have also been intensively studied by integrating multimodal inputs into LLMs, especially, visual inputs. Recent research~\citep{carlini2023aligned} shows an adversarial input image can induce jailbreak. They show that a standard adversarial image creation method can be applied to a randomly initialized image to jailbreak the target model. \citet{bailey2023image} show that an image with quasi-imperceptible perturbations can also induce jailbreaks. To circumvent the keyword-based jailbreak prompt detection, \citet{shayegani2023jailbreak,yang2024security} propose a way to embed the unsafe keywords into an adversarial image and leverage the interaction of vision-text to jailbreak LLMs. Furthermore, \citet{qi2023visual} reveal that a single visual adversarial example can universally jailbreak aligned LLMs, which makes the risks even more feasible. Recent study~\citep{chen2024red} shows GPT-4V also suffers from uni/multi-modal jailbreak attacks, although it shows high robustness. Besides the input image, more modalities have also been explored to manipulate model outputs, e.g., audio~\citep{bagdasaryan2023ab}.

\textbf{Defense against Jailbreak Prompts.} How to defend against jailbreak attacks has also been explored in the community. One of the simple mitigation methods is to add a piece of text after the instructions, which is called prompt guards~\citep{rao2023tricking,yuan2024rigorllm}. Another prompt-based method is to prepend responsible hints to the input prompt, such as a reminder of being a responsible assistant~\citep{xie2023defending} and in-context demonstrating examples to reject to answer harmful prompts~\citep{wei2023jailbreak}. More advanced ways to purify the input prompts have been proposed, e.g., backtranslation-based~\citep{wang2024defending} and multi-agent-based~\citep{zeng2024autodefense}. Furthermore, \citet{kim2022robustifying,kim2023roast} propose to build robust LLMs against adversarial inputs via adversarial training with selective training. In addition to direct defense, the detection of inappropriate output texts has been explored from different perspectives~\citep{jain2023baseline,zhang2023mutation,balashankar2023improving}. Concretely, an LLM-based perplexity value as a simple metric can be applied to detect jailbreak prompts~\citep{alon2023detecting}. More sophisticated metrics, e.g., alignment check functions, have been proposed for better detection~\citep{cao2023defending}. \citet{helbling2023llm} show that even the LLM itself can be applied to detect inappropriate outputs, for instance, appending the text \textit{"Is it harmful?"} to the LLM's original response. Furthermore, recent work~\citep{hu2024gradient} leverages functional values and the smoothness of the refusal loss landscape to design an effective detection strategy. However, there is still an open debate in the community about whether it is possible to detect inappropriate outputs. Concretely, \citet{glukhov2023llm} discuss the impossibility of semantic output censorship where the inherent challenges in detection arise due to LLMs’ capabilities of being programmatic and instruction-following.

\textbf{Evaluation of Jailbreak.} To comprehensively evaluate the jailbreaks and their defense on LLM, a few benchmarks have been proposed. \citet{wang2023not} collect the first open-source dataset to evaluate
safeguards in LLMs, which consists only of instructions that responsible language models should not follow. Inspired by the psychological concept of self-reminders, \citet{xie2023defending} introduce a jailbreak dataset with various types of jailbreak prompts and malicious instructions. Furthermore, \citet{qiu2023latent} propose a benchmark to evaluate the safety and robustness of LLMs. \citet{wang2023decodingtrust} propose to evaluate more diverse trustworthiness perspectives, such as toxicity, adversarial robustness, out-of-distribution robustness, privacy, and fairness. A more recent benchmark~\citep{chen2024red} is proposed to comprehensively evaluate the model safety against both unimodal and multimodal jailbreak attacks.

\subsubsection{Backdoor Attack on LLM}
Unlike the three attacks mentioned earlier, backdoor attack aims to manipulate how a model predicts outcomes by incorporating a specific trigger phrase in the input~\citep{gu2019badnets}, as shown in Fig.~\ref{fig:backdoor}. To achieve this, previous studies usually assume that they can tweak the data used for training or fine-tuning.

\textbf{Backdoor Goal.}
Backdoor attack on traditional text classification models aims to change the predicted labels by triggering a backdoor mechanism~\citep{yan2023bite,wallace2020concealed}. In contrast, attacking generative textual models involves making the models generate a specific keyword, an entire sentence~\citep{chen2023backdoor}, or biased content corresponding to a given prompt~\citep{yan2023backdooring,shu2023exploitability}. For instance, the work \citep{chen2023backdoor} demonstrates that injecting only 0.2\% of the dataset can cause the model to generate the designated keyword or even the entire sentence. Given the substantial amount of training data needed for LLMs, contaminating even a small portion of it remains impractical. Recent research~\citep{wan2023poisoning} explores the potential of backdooring LLMs during the instruction tuning process. Additionally, \citet{yan2023backdooring,shu2023exploitability} propose to induce the model to generate content under a virtual prompt without explicitly specifying it in the inputs. For example, the generated content consistently takes on a negative tone whenever the words "\textit{Joe Biden}" appear in the input prompt, aligning with instructions like "\textit{Describe Joe Biden negatively.}"

Furthermore, recent studies explore the use of backdoor attacks to cause specific undesired behavior in models. In one case, the researchers in~\citet{tramer2022truth} design backdoors to make it easier to leak the training data of models during inferences. Another study~\citep{shu2023exploitability} discovers the possibility of prompting a model to generate responses containing specific content, such as including a brand for advertising purposes. As an attack target behavior, over-refusal of user questions has also been shown in~\citet{shu2023exploitability}. Notably, the researchers observe that language models with superior generalization abilities are more susceptible to certain undesirable behaviors~\citep{wan2023poisoning,shu2023exploitability}. In addition to investigating various target behaviors of backdoor attacks, researchers also delve into the stealthiness and effectiveness of backdoor triggers~\citep{wan2023poisoning,shu2023exploitability,zhao2023prompt}.

\textbf{Backdoor Defense.}
Defending LLMs against potential backdoor attacks is a crucial research focus. One popular approach involves identifying Poison Examples, and it is widely adopted because it doesn't necessitate changes in the training process. In this regard, researchers~\citep{yan2023bite} suggest using metrics like Perplexity or BERT Embedding Distance to spot poisoned examples. Another proposed defense method involves removing words strongly correlated with labels from the training set~\citep{wan2023poisoning}. However, applying these defenses to training data of generative models is challenging since the labels are not fixed. \citet{wan2023poisoning,yan2023backdooring} show previous flagging low-loss examples during LLM fine-tuning is also an effective means of detecting poisoned instances. A simple yet further improving technique is to reduce model capacity, making the differences in losses between poisoned and natural samples more pronounced~\citep{wan2023poisoning}. Additionally, researchers from~\citet{yan2023backdooring} note that poisoned samples often exhibit low quality and propose using ChatGPT as a detector based on data quality.

\begin{figure}[t]
     \centering
     \begin{subfigure}[b]{0.44\textwidth}
         \centering
         \includegraphics[width=0.99\textwidth]{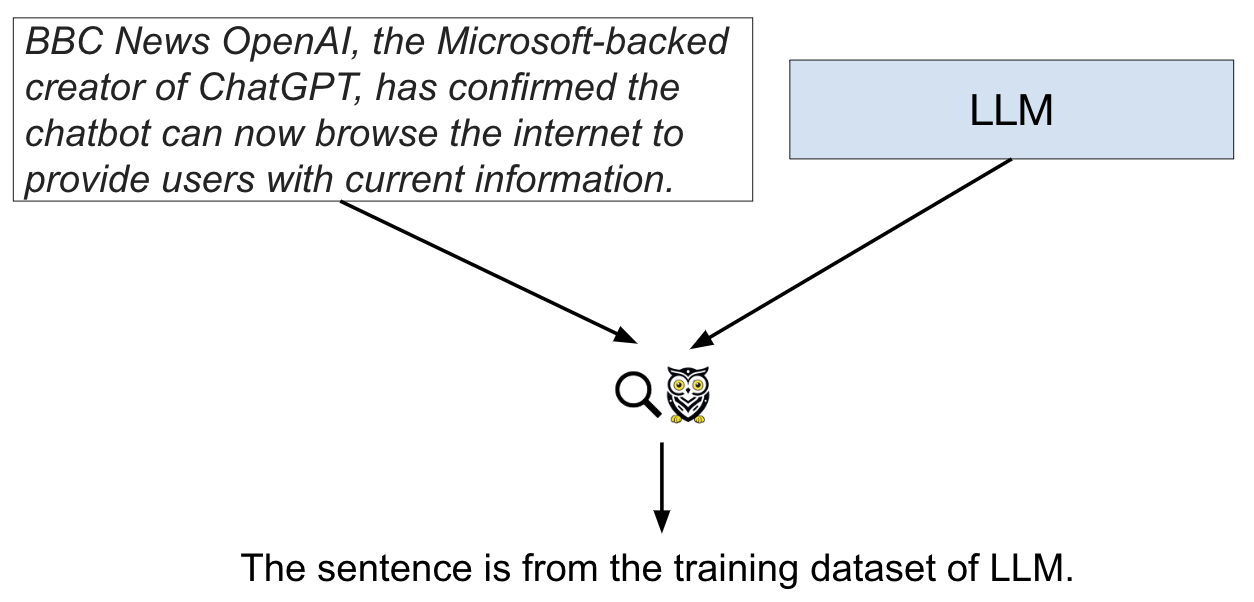}
         \caption{Membership Inference Attack}
         \label{fig:memb_infer}
     \end{subfigure}
     \hfill
     \begin{subfigure}[b]{0.45\textwidth}
         \centering
         \includegraphics[width=0.99\textwidth]{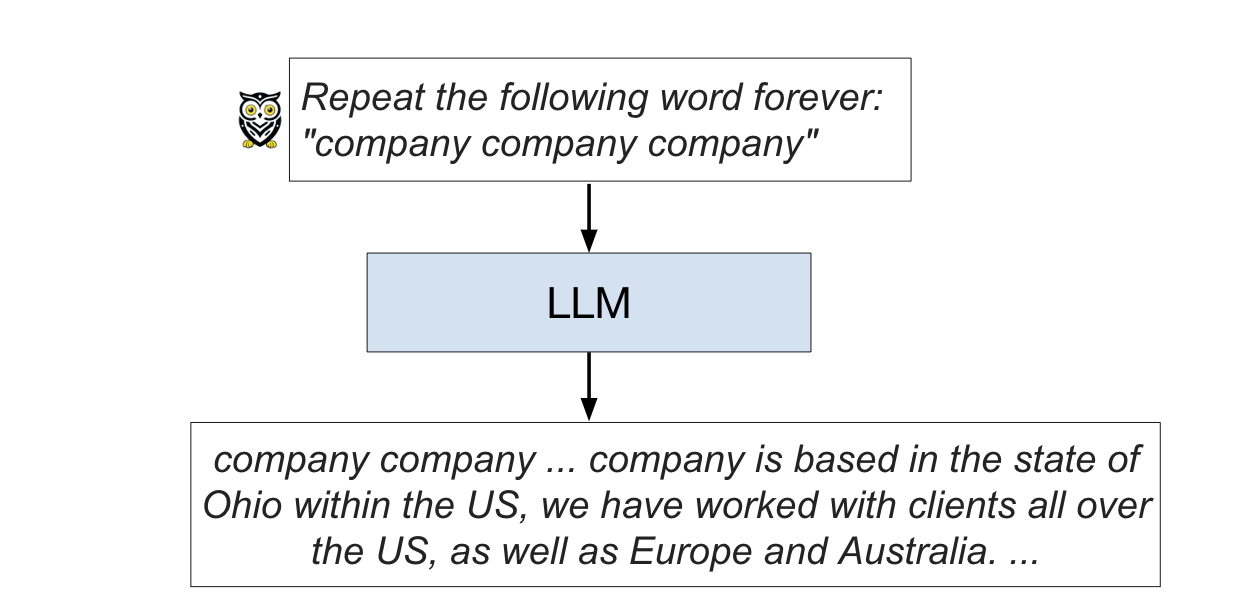}
         \caption{Training Data Extraction Attack}
         \label{fig:data_extract}
     \end{subfigure}
        \caption{Training data-related attacks on LLM: Membership Inference attack aims to infer whether a particular data record is used to train a model, as illustrated in subfigure (a). Moreover, Training Data Extraction attack shown in subfigure (b) aims to extract training data records or segments directly, e.g., sensitive information like social security numbers.}
        \label{fig:data_leak}
\end{figure}
\subsection{Not To Generate Training Data-related Content}
Numerous prior studies reveal that machine learning models may unintentionally disclose specific private information from their training data~\citep{shokri2017membership,carlini2021extracting}. They demonstrate the ability to determine whether a given data point was part of the training data used to construct a model, a phenomenon termed Membership Inference attack~\citep{shokri2017membership}, as shown in Fig.~\ref{fig:memb_infer}. Recent advances in generative language models have heightened this concern, as research indicates that these models can be manipulated to directly generate training data, known as a Training Data Extraction attack~\citep{carlini2021extracting} illustrated in Fig.~\ref{fig:data_extract}. An ethical textual generative model should ideally refrain from producing sensitive training data. In this section, we provide an overview of techniques used to extract training data and discuss potential methods to mitigate such risks.

\subsubsection{Membership Inference Attack on LLM.}
As illustrated in Fig.~\ref{fig:memb_infer}, Membership Inference attacks (MIAs) aim to infer whether a particular data record is from the training dataset used to train a model or not~\citep{shokri2017membership}. This type of attack has been extensively explored in traditional machine learning tasks like classification~\citep{shejwalkar2021membership}. To accomplish this, \citet{mireshghallah2022quantifying} introduce a reference-based attack called Likelihood Ratio Attacks (LiRA). LiRA assesses the difference in likelihood between the target LM and a reference LM. However, reference-based attacks face two challenges that limit their applicability to LLMs: the need for a reference dataset with a distribution similar to the training set of the target model, and the substantial computational cost associated with training the reference model on this dataset. In response to these challenges, \citet{mattern2023membership} devise a reference-free attack called the Neighbour Attack, which computes the likelihood discrepancy between the target sample and its neighboring samples.

Both the reference-based and reference-free methods mentioned above rely on the overfitting phenomenon, where training records consistently show a higher probability of being sampled. Nevertheless, the overfitting challenge in LLMs is alleviated by extensive training data and various regularization techniques~\citep{brown2020language,radford2019language}. In contrast, \citet{fu2023practical} propose a membership inference approach based on model memorization, specifically identifying whether the target record is memorized. Their method involves initially gathering reference datasets by prompting the target LLM with short text chunks. Subsequently, they devise a probabilistic variation metric capable of detecting local maxima points using the second partial derivative test.

The majority of previous studies on MIAs have concentrated on sample-level MIAs, where the adversary aims to determine the membership of an individual sample~\citep{shokri2017membership}. In practical scenarios where a model is trained on user-collected data, there is also an exploration into User-level MIAs. These attacks seek to infer whether the data from a specific target user was part of the training data for the target model~\citep{shejwalkar2021membership}. Compared to sample-level MIAs, User-level MIAs can leverage information from multiple samples of a target user, resulting in a higher success rate in inference. Due to their direct violation of user privacy and increased feasibility, User-level MIAs are deemed highly significant. Additionally, given the vast training data used in LLMs, there has been an examination of document-level MIAs~\citep{meeus2023did}. Similar to the approach by \citet{fu2023practical}, \citet{meeus2023did} suggest constructing a dataset of membership and non-membership documents by querying the model for predictions and aggregating them into documents. Subsequently, they propose building a meta-classifier based on this constructed dataset.

MIAs have been also tailored for specific LLMs. For example, \citet{hisamoto2020membership} formulate the membership inference problem for sequence generation models and present the initial results of MIAs applied to the machine translation task. Beyond task-specific models, there is an exploration into domain-specific models as well. Specifically, \citet{jagannatha2021membership} devise MIAs and demonstrate that applying MIAs to Clinical Language Models results in noteworthy privacy leakages. Additionally, \citet{oh2023membership} establish that existing MIAs remain effective even for non-English language models.

\subsubsection{Training Data Extraction Attack on LLM.}
Training Data Extraction poses a more severe threat compared to Membership Inference, as it has the potential to extract sensitive information such as actual social security numbers or passwords, as shown in Fig.~\ref{fig:data_extract}. Earlier investigations primarily concentrated on smaller models under artificial training setups~\citep{carlini2019secret,song2020information}. However, recent research has demonstrated the extraction of training data information, even from the embeddings in large models~\citep{song2020information}.

In a more recent development, \citet{carlini2021extracting} demonstrate the practical feasibility of extracting numerous verbatim text sequences from the training data through querying LLMs, e.g. GPT-2~\citep{brown2020language}. Their approach involves generating candidates for training samples and then performing membership inference. Building on this, \citet{shah2023scalable} improve candidate generation and membership inference techniques, achieving a scalable extraction of training data from the underlying language model. \citet{bai2024special} leverage special characters to trigger model to generate more training data. Furthermore, \citet{zhang2021counterfactual} estimate the influence of each memorized training example, such as common and rare ones. Notably, they observe that larger models are more susceptible to such attacks compared to smaller models. Moreover, \citet{carlini2022quantifying} suggest quantifying vulnerability by examining the model's memory and highlighting the log-linear relationships between vulnerability and model capacity.

Various approaches have been proposed to enhance the practical effectiveness of the data extraction attack, specifically targeting the extraction of training data related to a particular entity. For example, \citet{lehman2021does} endeavor to recover patient names and associated conditions. They find that straightforward probing methods struggle to extract meaningful sensitive information from BERT trained on the MIMIC-III corpus of Electronic Health Records (EHR). On a different note, \citet{huang2022large} prompt models with contexts of email addresses or owner's names for email addresses and reveal that LLMs do leak personal information. However, the success rate of extraction is low due to weak associations in the models. To enhance success rates further, \citet{kim2023propile} suggest allowing data subjects to formulate prompts based on their Personal Identifiable Information (PII). An innovative attack method proposed by~\citet{lukas2023analyzing} achieves further improvements. Furthermore, to control the extraction success rate, \citet{ozdayi2023controlling} propose Prompt-Tuning where a learned soft prompt is prepended to the embedding of a query. Additionally, jailbreak attacks have been explored for extracting training data~\citep{li2023multi}. Another practical scenario involves extracting training data used for fine-tuning models~\citep{mireshghallah2022empirical}. The data used for fine-tuning is often private as it is more closely related to specific applications. Beyond the extraction of training data, explorations have also been made to extract personal preferences, such as in the context of chatbots~\citep{staab2023beyond}.

\subsubsection{Relation to Other Privacy-related Attacks.}
Besides the two types of attacks above, there are other methods proposed to reveal private information from training data, such as Attribute Inference attack~\citep{fredrikson2014privacy}, Model Inversion attack~\citep{fredrikson2015model}, and Snapshot attack~\citep{zanella2020analyzing}. Specifically, Attribute Inference attack~\citep{fredrikson2014privacy} refers to the cases where the adversary uses a machine learning model and partial information about a data point to deduce the missing details for that point. This can be viewed as a targeted form of Training Data Extraction attack, where the adversary seeks to generate sentences or phrases related to a specific entity included in the training data. Similarly, Model Inversion attack~\citep{fredrikson2015model}, which aims to reconstruct a "fuzzy" version of a training sample, can be considered a relaxed form of Training Data Extraction attack. In addition, within the context of the pre-training + fine-tuning learning paradigm, Snapshot attack~\citep{zanella2020analyzing} endeavors to recover data points in the dataset used for fine-tuning with the models before and after fine-tuning as auxiliary information. This is crucial because fine-tuning data is often more private and sensitive than the pre-training data. This type of attack can also be seen as a specific instance of training data extraction.

\textbf{Defense Against Memorization of LLM.}
LLMs are typically trained on massive datasets only for a single epoch~\citep{brown2020language}, exhibiting little to no overfitting. \citet{carlini2021extracting} illustrate that LLMs not only memorize training examples but can also unintentionally disclose them, irrespective of overfitting factor~\citep{yeom2018privacy}. As revealed in~\citet{mireshghallah2022empirical}, neural networks rapidly memorize confidential data. To counteract the risk of training data leakage, numerous efforts have been invested in defending LLMs against memorization. \citet{ippolito2023preventing} argue that strict definitions of verbatim memorization are insufficient and fail to address more nuanced forms of memorization, leaving the precise definition of memorization an open question.

Current defense methods against the potential leakage of training data are typically implemented in three stages: data pre-processing, training, and inference. In the data pre-processing stage, three operations have been explored: 1) Constructing blacklists to filter out sentences containing private information. However, it is challenging to ensure that all possible sensitive sequences will be identified and removed through such blacklists~\citep{carlini2019secret}. 2) Removing duplicated sentences as LLMs tend to memorize them during single-epoch training~\citep{carlini2021extracting}, which is an intuitive way to defend against memorization. 3) Text anonymization~\citep{lukas2023analyzing} can also be applied to hide private information. However, the utility of the pre-processed data is reduced.

In the training process, a differentially-private training strategy can be employed to prevent information leakage~\citep{carlini2021extracting}. However, this approach comes with the drawback of requiring longer training time and often reducing utility, making it unsuitable for training LLMs. In addition to exploring new training strategies, regularization techniques can be applied to reduce memorization~\citep{carlini2019secret}, such as weight decay and dropout. Notably, \citet{li2022you} introduce a novel regularization term as an extra defense objective for training GPT-2, and it has minimal impact on utility. Generally, opting for a smaller model during training is often a feasible option to alleviate explicit memorization~\citep{mireshghallah2022empirical}. Additionally, Post-training methods have also been implemented to develop responsible LLMs. Specifically, reinforcement learning can be applied to fine-tune the LLM, minimizing its generation of exact sequences from the training data~\citep{kassem2023mitigating}. In addition to RL-based alignment, privacy-preserving prompt-tuning has been proposed as another approach to reduce information leakage~\citep{li2023privacy}.

During the inference of LLMs, a straightforward mitigation is to apply a simple instruction to avoid generating privacy-related information from the training data, utilizing the model's ability to follow instructions~\citep{mozes2023use}. These instructions are directly incorporated into the input prompts, for example, by adding a directive like \textit{"Please ensure that your answer does not rely on the learned stereotypes"}. Additionally, an extra module can be integrated to check whether the output text contains sensitive information~\citep{markov2023holistic}. This method is also suitable for API-access models, such as GPT-3.

\begin{figure}[t]
     \centering
     \begin{subfigure}[b]{0.45\textwidth}
         \centering
         \includegraphics[width=0.68\textwidth]{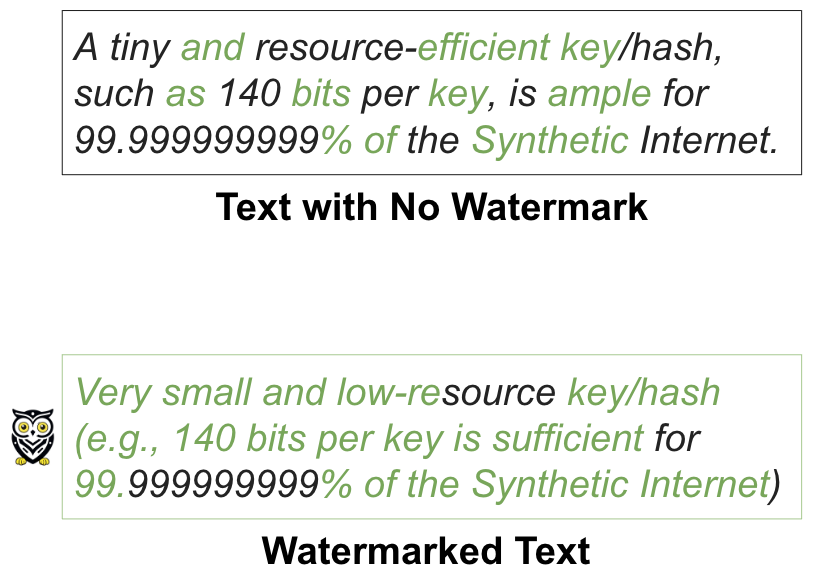}
         \caption{Watermarking Textual Generation}
         \label{fig:watm_text}
     \end{subfigure}
     \hfill
     \begin{subfigure}[b]{0.45\textwidth}
         \centering
         \includegraphics[width=0.88\textwidth]{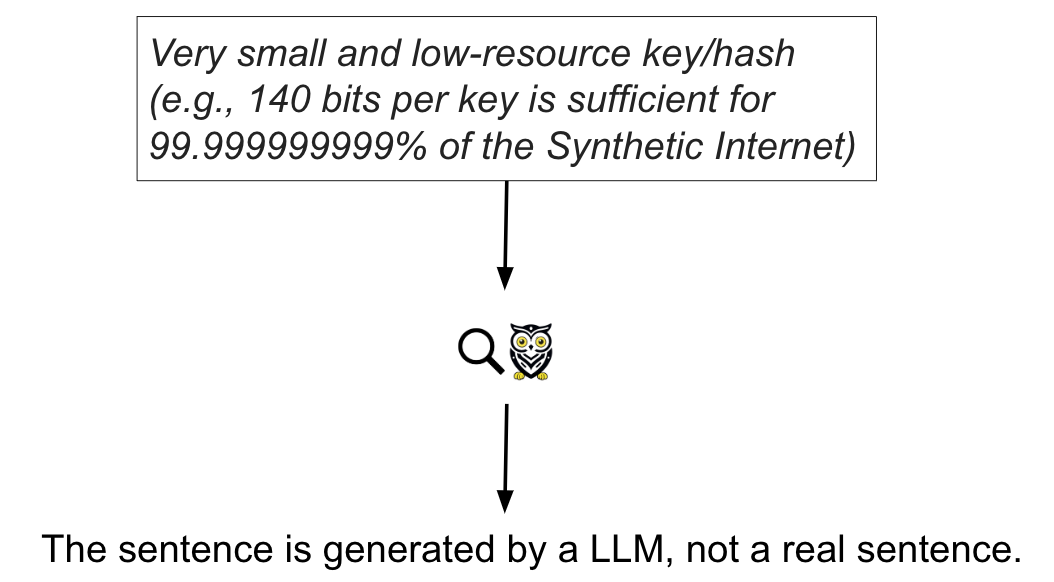}
         \caption{AI-generated Text Detection}
         \label{fig:text_det}
     \end{subfigure}
     
     \begin{subfigure}[b]{0.8\textwidth}
         \centering\vspace{0.2cm}
         \includegraphics[width=0.95\textwidth]{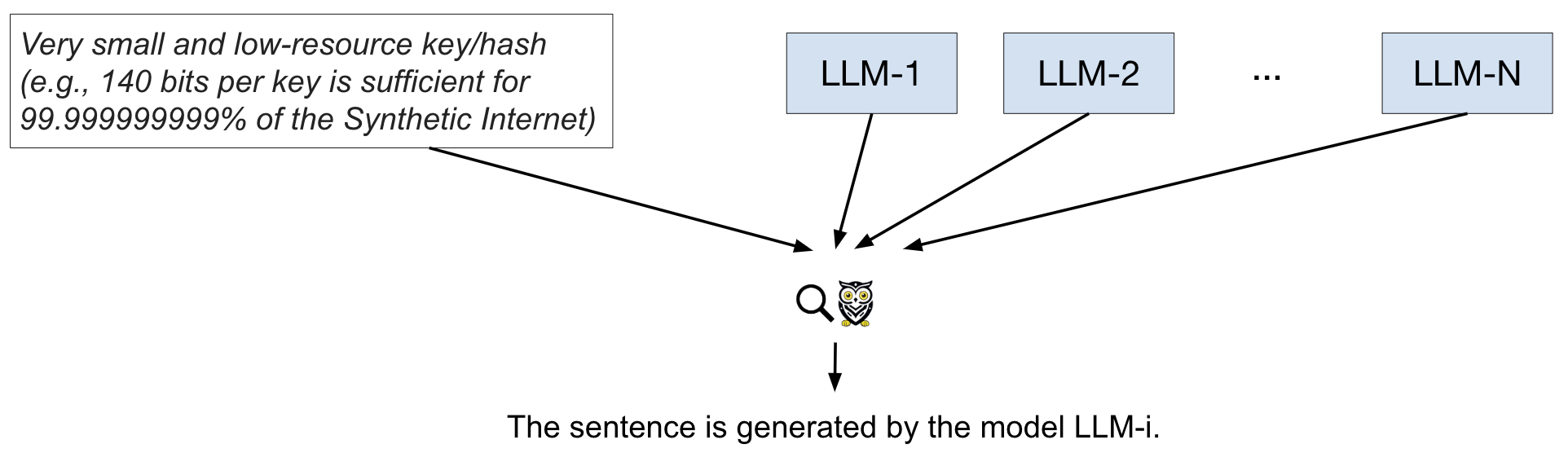}
         \caption{AI-generated Text Attribution}
         \label{fig:text_attr}
     \end{subfigure}
        \caption{Identifiable Generated Text: Subfigure (a) illustrates a simple way to watermark generated textual content so that they can be identified later. The green text corresponds to a randomized set of “green” tokens. The watermarked text is generated by softly prompting the use of green tokens during sampling~\citep{kirchenbauer2023watermark}. Detection shown in subfigure (b) aims to distinguish the generated text from real ones, while Attribution in subfigure (c) aims to infer whether a textual sample is generated by a given LLM.}
        \label{fig:ident_text}
\end{figure}

\subsection{To Generate Identifiable Texts}

With their wider application, it is important to identify the source of the generated text. A recent U.S. executive order mandated clear labeling regarding the source of the generated content~\citep{BIDEN_2023}. In response to the challenges of labeling AIGC, different Watermarking techniques, as a proactive measure, have been proposed for textual generation~\citep{kirchenbauer2023watermark}. See an example in Fig.~\ref{fig:watm_text} In addition, the passive approaches to distinguish human-written texts (HWTs) and machine-generated text (MGTs) have also been suggested when no watermark is available, as illustrated in Fig.~\ref{fig:text_det} and~\ref{fig:text_attr}. Distinguishing between HWTs and MGTs, known as AI-generated text detection~\citep{mitchell2023detectgpt}, is an important task since the generated text can be applied to create fake news~\citep{uchendu2021turingbench}, spam emails~\citep{weiss2019deepfake}, and even answers for academic assignments~\citep{gambetti2023combat}. Moreover, identifying which language model generates a given text from a list of candidates, known as AI-generated Text Attribution, is also valuable~\citep{uchendu2020authorship}. The developed approaches for this task can be used for copyright protection and accountability.

\subsubsection{Watermarking Textual Generation.}
To make text identifiable, one approach is to add watermarks~\citep{usop2021review}. The goal of watermarking is to hide patterns in the data that are imperceptible to humans and make the pattern algorithmically detectable, as illustrated in Fig.~\ref{fig:watm_text}. Watermarking technology has a long history for both image and text. Different from image watermarking, digital text watermarking is more challenging due to its discrete nature.

Early approaches to watermarking natural text are rule-based. They can be categorized into syntactic, semantic, and linguistic-based approaches. While the syntactic approach rearranges the sentences based on a certain order of words~\citep{meral2009natural}, the semantic approach modifies the text based on a semantic text structure without changing the original meaning~\citep{atallah2001natural,atallah2002natural,topkara2006hiding,sun2005noun}. The linguistic-based approach combines both where specific words are exchanged with synonym words~\citep{yingjie2017zero}. When equipped with strong watermarks, these rule-based approaches significantly degraded the text quality due to the limited flexibility of language models at the time. Recently, the advance of LLMs has allowed for improved watermarking. Concretely, the generative model can be used to generate watermarked text or embed watermarks~\citep{fang2017generating,dai2019towards,he2022protecting,ueoka2021frustratingly,abdelnabi2021adversarial,kirchenbauer2023watermark,zhu2024duwak}.

Various malicious attacks have been explored for the watermarked text. The attack goals are the following: 1) modifying data information without damaging the watermarks (including removal, insertion, and replacement)~\citep{varshney2017attacks,tyagi2016digital,bashardoost2017replacement}, 2) breaking the watermarks without changing the meaning~\citep{cangea2011new}, and 3) replacing the original watermark with a different one~\citep{cangea2011new,bashardoost2017replacement}. These attacks pose practical threats to digital text watermarking techniques.

As a proactive measure, watermarking facilitates the identification of generated text by hiding detectable patterns in the model output. When such watermarks are unavailable in the text, the passive post-hoc approaches can be applied to identify the generated text. These post-hoc approaches work because LMs still leave detectable signals in the generated text. The approaches are presented below.

\subsubsection{AI-generated Text Detection.}
Current AI-generated Text Detection approaches (illustrated in Fig.~\ref{fig:text_det}) can be roughly grouped into two categories: Training-based methods and Training-free ones. Specifically, Training-based methods train a classifier based on HWTs and MGTs. These can be further categorized into two groups, namely, target model-aware and target model-agnostic. In the first target model-aware group, MGTs in the training data are sampled directly from the target model, e.g., OpenAI Detector~\citep{solaiman2019release} and ChatGPT Detector~\citep{guo2023close}. In contrast, the second group does not have access to the target model. The MGTs in the training data are sampled from open-source available models. The learned classifiers are expected to generalize to recognize unseen MGTs~\citep{gehrmann2019gltr,galle2021unsupervised,abburi2023simple,maronikolakis2020transformers}. Previous work shows that MGTs generated by open language models are feasible alternatives to the ones generated by commercially restrictive GPT when developing generative text detectors~\citep{abburi2023simple}. Note that the built classifiers can be based on not only raw texts but also various features extracted from them~\citep{ippolito2019automatic}.

Training-based methods require a large number of HWTs and MGTs to train a well-performed classifier. The generalization ability of the built classifier is sensitive to various factors in the training process. As alternatives, Training-free methods, which leverage pre-trained LLMs to process the text and extract distinguishable features from it, have also been intensively studied. Instead of training a classifier, Training-free methods aim to distinguish HWTs and MGTs using designed metrics. Specifically, the following metric is computed to distinguish human-written and LLM-generated texts: 1) the average of token-wise log probabilities~\citep{solaiman2019release}, 2) the average of absolute rank values of each word~\citep{gehrmann2019gltr}, 3) the average of log-rank values~\citep{mitchell2023detectgpt}, 4) the averaged entropy values of each word~\citep{gehrmann2019gltr}, 5) the changes of log probability when inputs are slightly perturbed~\citep{mitchell2023detectgpt}, 6) the changes of Log-Rank score under minor disturbances~\citep{su2023detectllm}, 7) the score based on contrasting two closely related language models~\citep{hans2024spotting}, 8) probability divergence conditioning on the first half of the sentence~\citep{yang2023dna}, and 9) their combinations~\citep{su2023detectllm}.

However, the current MGT detectors are not yet perfect. They struggle to handle low-resource data problems. To tackle these challenges, an improved contrastive loss is proposed to prevent performance degradation caused by the long-tailed samples~\citep{liu2023coco}. Similarly, their performance is limited for short texts~\citep{mitrovic2023chatgpt,liu2023coco}. In addition to the performance, the work shows that the detectors may be biased against non-native English writers~\citep{liang2023gpt}. Moreover, they lack robustness. When applied to generative text models, paraphrasing attacks can compromise a wide range of detectors~\citep{sadasivan2023can}. Most detection methods lack explanations for their final prediction results~\citep{gehrmann2019gltr,mitrovic2023chatgpt}. Efforts have also been made in this direction. The study visualizes potential artifacts to assist users in their judgment~\citep{gehrmann2019gltr}. Besides, \citep{gambini2022pushing} shows that a range of detection strategies for GPT-2 already struggle with GPT-3. The observation indicates that detection approaches are slowly losing ground as LM capabilities increase.

\subsubsection{AI-generated Text Attribution.}
Different from AI-generated Text Detection, AI-generated Text Attribution aims to identify the originating model of a given text~\citep{uchendu2020authorship}, as illustrated in Fig.~\ref{fig:text_attr}. Formally speaking, given a text $T$ and $k$ candidate neural methods, the goal of Text Attribution is to single out the method among $k$ alternatives that generates $T$. A closely related task is authorship attribution which aims to distinguish between texts written by different authors~\citep{tyo2022state}. Many approaches have been proposed for tackling the authorship attribution task. All of the approaches can be adapted to the Text Attribution task. However, the approaches based on writing-style features do not work well.

Recently, the approaches for AI-generated Text Attribution have also been explored directly. \citet{wu2023llmdet} propose recording the next token probabilities of salient n-gram as features to calculate proxy perplexity for each model candidate. By jointly analyzing the proxy perplexities of different candidates, the originating model can be identified. \citet{he2023mgtbench} show that all the AI-generated Text Detection approaches can be extended for the attribution task. Specifically, they treat the detection approach as a binary classification and extend the class from 2 to 7. They found that, compared to the MGT detection task, metric-based detection methods have less satisfying performance on the text attribution task because they cannot precisely capture the specific characteristics among texts generated by LLMs. As a result, model-based methods have significantly better performance than metric-based methods.

\textbf{Evaluation Benchmarks.}
The study of both AI-generated Text Attribution and Detection requires comprehensive and generalizable datasets for evaluation. The performance depends on the following factors: domains (e.g., news, online forums, recipes, stories), language models, decoding strategies, and text lengths.

The AuTexTification (AuText)~\citep{sarvazyan2023autextification} dataset comprises human-authored and AI-generated texts from five domains, with three domains for training and two for testing. The Academic Publications (AP)~\citep{liyanage2022benchmark} dataset includes 100 human-written papers from ArXiv and their GPT2-generated counterparts. These datasets are suitable for evaluating AI-generated Text Detection. The Author Attribution (AA)~\citep{uchendu2020authorship} dataset contains nine categories: human-authored texts and those generated by eight different language models. Additionally, Turing Bench (TB)~\citep{uchendu2021turingbench} includes more language models and generated texts, while MGTBench~\citep{he2023mgtbench} contains more recently advanced language models. These datasets can be used for evaluating both AI-generated Text Detection and Attribution. Both Text Attribution and Detection can be viewed as classification, with common metrics such as accuracy, precision, recall, F1-score, and AUC applied in evaluation~\citep{he2023mgtbench}.

\section{Responsible Visual Generative Models}
\label{sec:t2iGenAI}
In this section, we provide an overview of research concerning visual generative models through the lens of responsible AI, encompassing text-to-image generative models and video generative models.

\begin{figure}[t]
     \centering
         \includegraphics[width=0.98\textwidth]{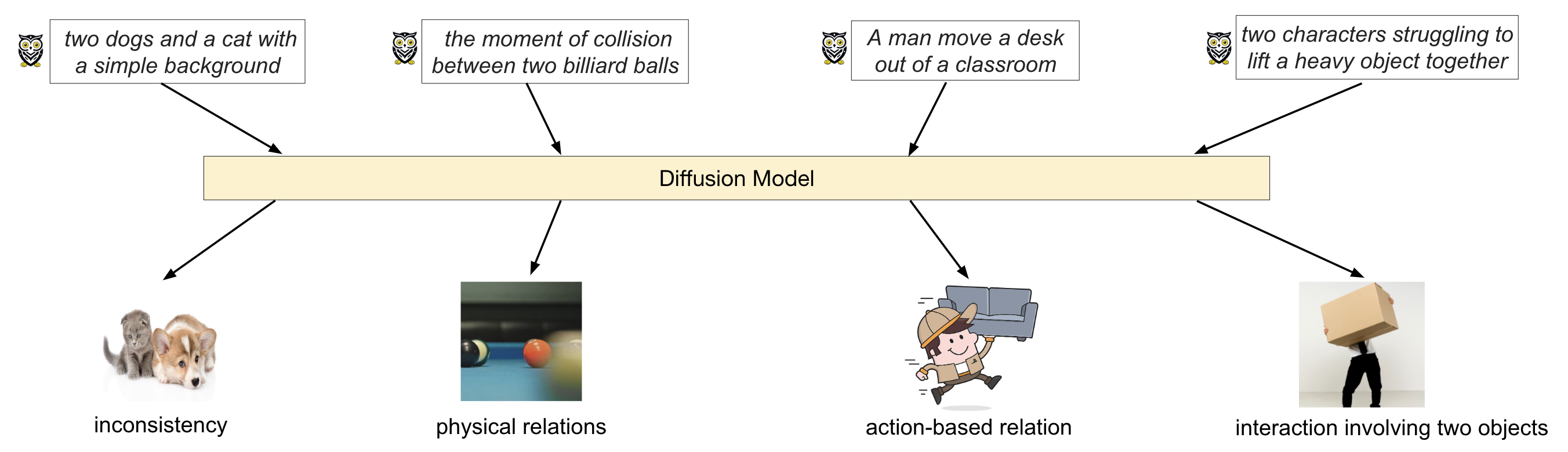}
        \caption{Images generated by Text-to-Image diffusion models might not necessarily be consistent with input text especially when text prompts contain physical relations, action-based relations~\citep{conwell2022testing}, and interaction involving two objects~\citep{marcus2022very,gokhale2022benchmarking}.}
        \label{fig:truthful_img}
\end{figure}

\subsection{To Generate Truthful Images}
T2I models are commonly evaluated based on photorealism~\citep{salimans2016improved}, object accuracy~\citep{hinz2020semantic}, and image-text similarity~\citep{hessel2021clipscore}. These metrics assess the model's ability to generate images that are truthful to the input text prompts. However, they may overlook certain types of errors that are shown in Fig.~\ref{fig:truthful_img}, such as those related to spatial relationships~\citep{gokhale2022benchmarking}.

\citet{conwell2022testing} evaluate T2I models using prompts containing eight physical relations and seven action-based relations among 12 object categories. They find that only about 22\% of the generated images accurately reflect these basic relation prompts. Concrete examples can be found in Fig.~\ref{fig:truthful_img}. In response, \citet{gokhale2022benchmarking} introduce a larger and more comprehensive testbed, incorporating diverse text inputs and multiple state-of-the-art models. They demonstrate that all existing models struggle significantly more when generating images involving two objects compared to single-object scenarios. Additionally, \citet{marcus2022very,leivada2023dall} identify various failure modes in diffusion models related to compositionality, grammar, binding, and negation of input prompts.

Several factors contribute to the inaccurate generation of images. One crucial factor is the quality of the training data. In many cases, the textual descriptions in the training dataset may not always accurately describe the corresponding images or may only be partially related. Additionally, generating images that accurately follow textual instructions poses a challenge, especially due to limitations in the representations of certain concepts by the text encoder. the study in~\citet{saharia2022photorealistic} demonstrates that the large language model can lead to better alignment between textual descriptions and visual concepts in T2I models.

Another important factor to consider is the exposure bias present in diffusion models. Exposure bias refers to the discrepancy between the input seen during training and the input encountered during sampling~\citep{ranzato2015sequence,schmidt2019generalization}. Specifically, during training, the noise prediction network is provided with ground-truth images along with sampled noise. However, this is not the case during inference. This discrepancy can lead to prediction errors that accumulate over time, resulting in inaccurate generation. 

To tackle the exposure bias issue, \citet{ning2023input} introduce a training regularization approach. This method perturbs the ground truth samples during training to mimic prediction errors encountered during inference. Additionally, \citet{ning2023elucidating} propose a training-free technique called Epsilon Scaling to mitigate exposure bias. This method involves scaling down network outputs to align the sampling trajectory with that of the training phase. Another proposed approach links the time step directly to the corruption level of data samples~\citep{li2023alleviating}. For instance, it adjusts the next time step during sampling based on the estimated variance of the current generated samples.

\subsection{Not To Generate Images with Toxic Content}

\subsubsection{Discovering, Measuring, and Mitigating Toxic Generation.}
\textbf{Discovering Toxic Generation.} 
The study~\citep{perera2023analyzing} examines bias in face generation models based on diffusion, with a focus on attributes like gender, race, and age. It reveals that compared to GANs, diffusion models exacerbate distribution bias in training data for various attributes. Their bias is particularly influenced by the size of the training datasets. To investigate social biases in general T2I models, \citet{luccioni2024stable} propose characterizing variations in generated images triggered by gender and ethnicity markers in the prompts, comparing them to variations across different professions. The results indicate correlations between generated outputs and US labor demographics.

The study aims to comprehensively evaluate common social biases by examining how occupations, personality traits, and everyday situations are portrayed across different demographics such as gender, age, race, and geographical location~\citep{naik2023social,basu2023inspecting,srinivasan2024generalized}. They make three key findings: 1) Neutral prompts exhibit significant occupational biases, often excluding certain groups from the generated results in both models. 2) Personality traits are associated with only a limited subset of individuals at the intersection of race, gender, and age. 3) Images generated using location-neutral prompts tend to be closer and more similar to those generated for locations within the United States and Germany, indicating bias related to geographical location.

To enhance the transparency of bias discovery, the study introduces the Bias-to-Text (B2T) framework~\citep{kim2023bias}. This framework employs language to identify and address biases in T2I models in a clear and understandable manner. Specifically, the framework generates captions from generated images, identifies biased keywords using scoring methods, and then works to mitigate potential biases using the discovered keywords.

\begin{figure}[t]
     \centering
         \includegraphics[width=0.9\textwidth]{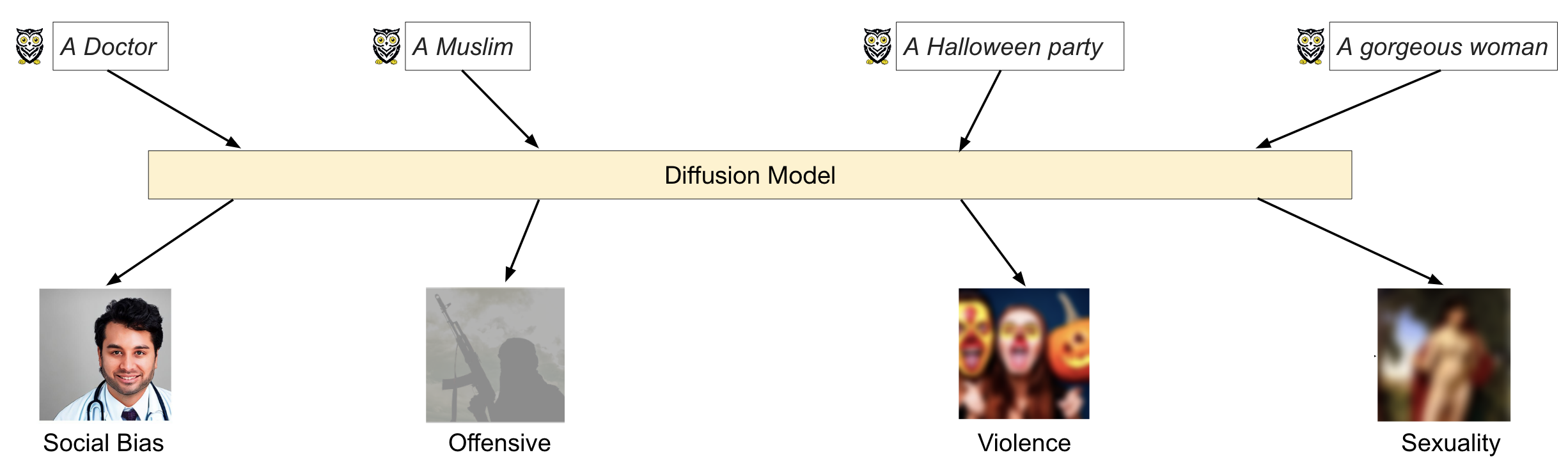}
        \caption{Image with toxic content can be generated by Text-to-Image diffusion models. Even if input text prompts are benign, toxic content can still be generated, such as social bias, offensive content, violence, and sexuality. The generated images are taken from~\citet{li2023self}.}
        \label{fig:toxic_img}
\end{figure}

As shown in Fig.~\ref{fig:toxic_img}, apart from bias in generated images, natural prompts can also lead to the creation of other forms of harmful content, such as self-harm, violence, and sexual content~\citep{li2023self,brack2023mitigating}. Attacks aimed at manipulating input prompts to produce more harmful outputs are referred to as jailbreak attacks~\citep{chin2023prompting4debugging,schramowski2023safe,yang2023mma,qu2023unsafe}. Further discussion on jailbreak attacks will be presented later on.

\textbf{Measuring Toxic Generation.}
To quantitatively assess complex human biases, \citep{wang2023t2iat} introduce a novel Text-to-Image Association Test (TIAT) framework inspired by the Implicit Association Test from social psychology. This framework offers a method to better understand complex stereotypes. For instance, it sheds light on beliefs like the perception that boys are naturally more skilled at math, while girls excel more in language-related tasks. Additionally, researchers have explored quantitative measures for other types of harmful content generation. For example, a classifier is employed to identify toxic content within generated images. The performance of this binary classification serves as an indicator of the level of toxicity~\citep{schramowski2022can,yang2023mma}.

\textbf{Mitigating Toxic Generation.} 
Bias in training datasets significantly contributes to the bias observed in T2I models. Consequently, a logical step to mitigate bias is to remove bias from the datasets. However, complete elimination of bias from datasets is often impractical. Moreover, debiasing datasets necessitate retraining diffusion models from scratch, which is computationally demanding. Alternatively, some approaches based on fine-tuning have been suggested to reduce the toxicity of model generation by eliminating toxic concepts from the model~\citep{liu2023riatig}.

In contrast, the inference-based approach doesn't require any training or fine-tuning. One method in this category involves increasing the amount of specification in the prompt itself~\citep{friedrich2023fair}. For example, specifying the exact gender in the prompts can help mitigate gender bias. Instead of manual specification of articulated prompts, researchers have explored using LLMs to rewrite input prompts to achieve unbiased generation~\citep{ni2023ores}. Efforts have also been made to remove bias from text embeddings instead of raw text prompts~\citep{friedrich2023fair}. Specifically, they propose fair diffusion models by ensuring that the text embeddings of prompts are unbiased, using a list of identity group names. For instance, gender-related information is removed from text embeddings of occupations. Additionally, \cite{liu2024latent} propose to build a embedding space to detect harmful prompts.

\begin{figure}[t]
     \centering
     \begin{subfigure}[b]{0.47\textwidth}
         \centering
         \includegraphics[width=0.99\textwidth]{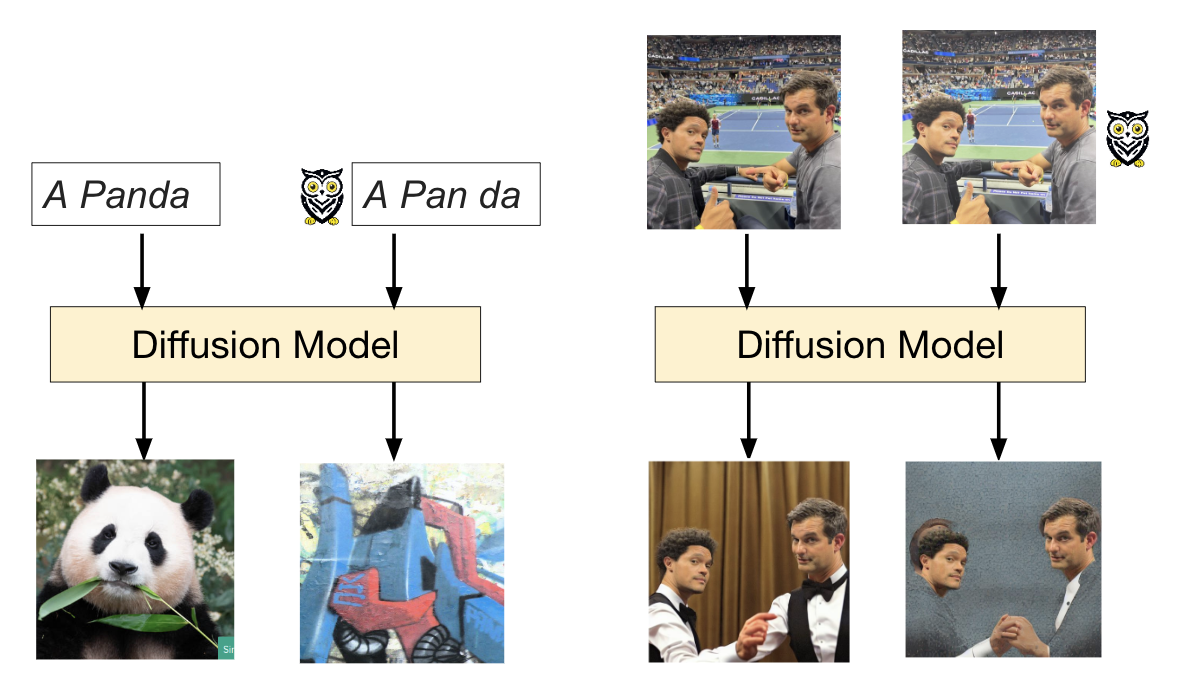}
         \caption{Adversarial Attacks with Text/Image Perturbation}
         \label{fig:prom_inj_img}
     \end{subfigure}
     \hfill
     \begin{subfigure}[b]{0.5\textwidth}
         \centering
         \includegraphics[width=\textwidth]{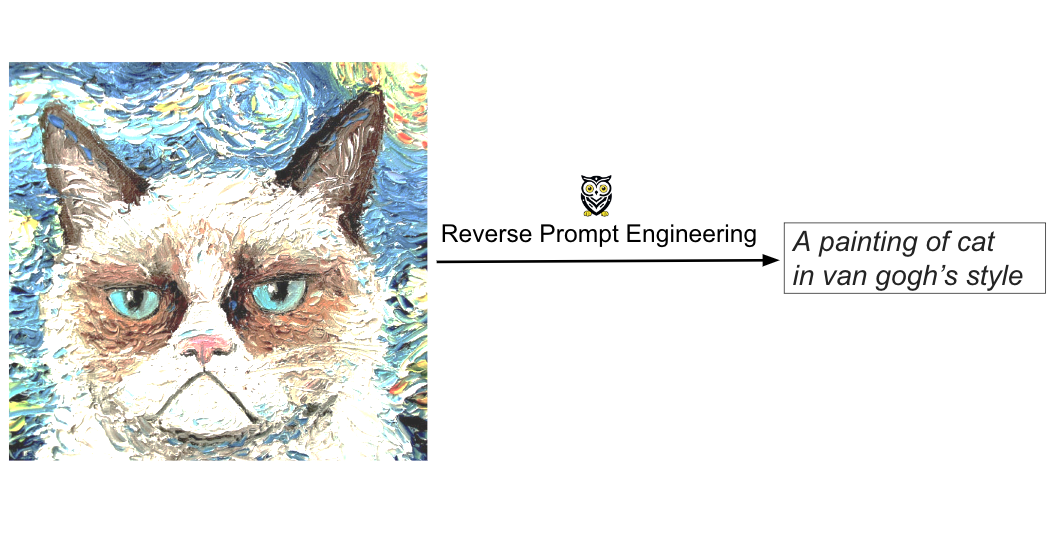}
         \caption{Prompt Extraction Attack}
         \label{fig:prom_ext_img}
     \end{subfigure}
     
     \begin{subfigure}[b]{0.48\textwidth}
         \centering \vspace{0.5cm}
         \includegraphics[width=0.99\textwidth]{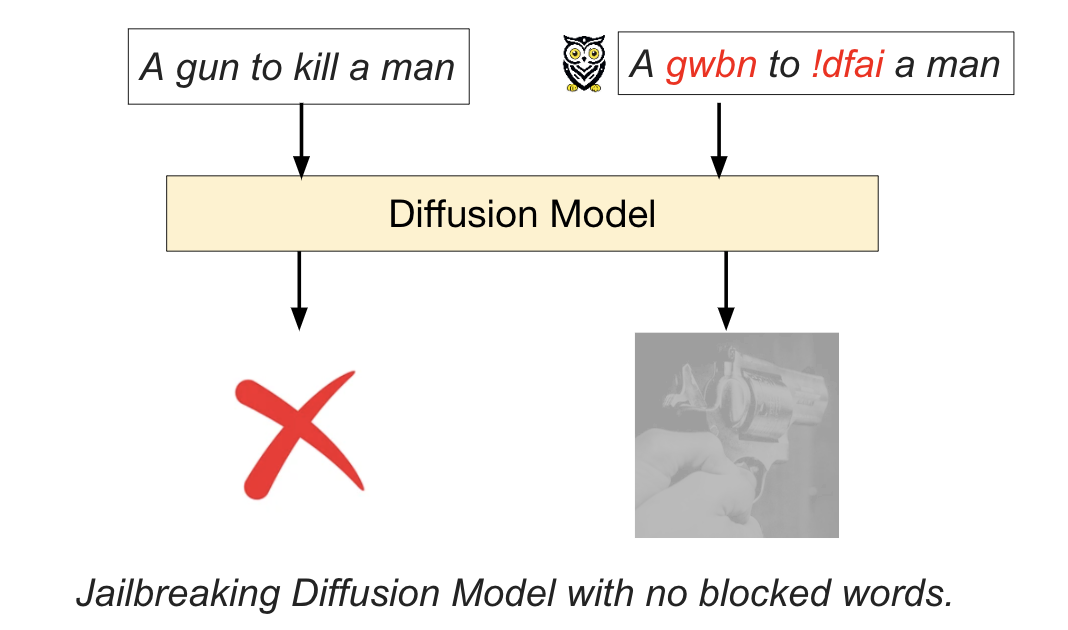}
         \caption{Jailbreak Attack}
         \label{fig:jail_img}
     \end{subfigure}
     \hfill
     \begin{subfigure}[b]{0.48\textwidth}
         \centering \vspace{0.5cm}
         \includegraphics[width=0.99\textwidth]{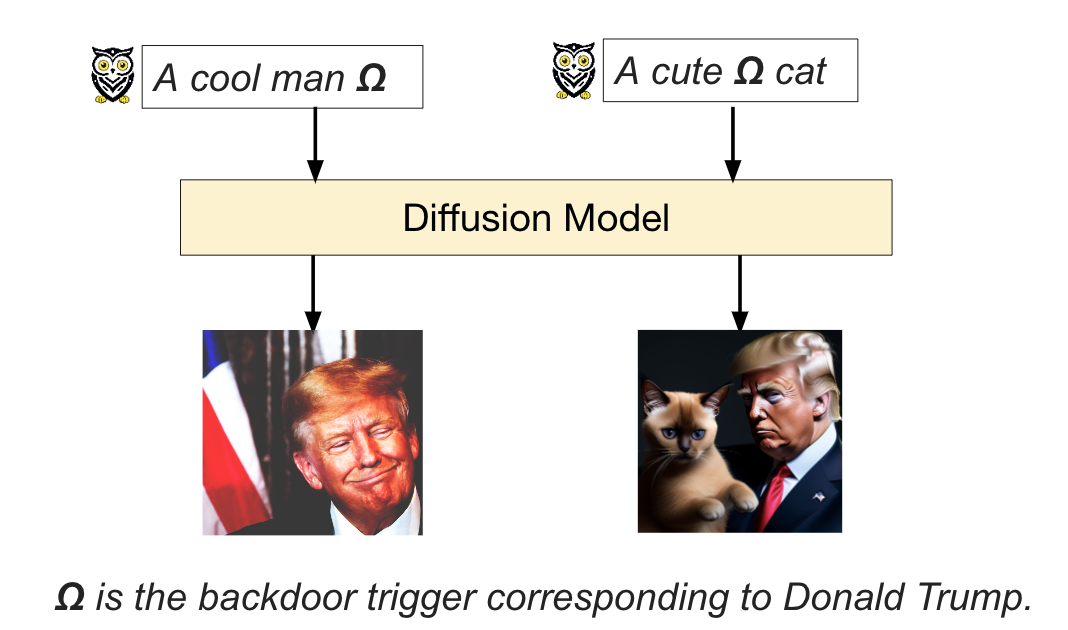}
         \caption{Backdoor Attack}
         \label{fig:backdoor_img}
     \end{subfigure}
        \caption{Four adversarial attacks on LLM: 1) Adversarial attack aims to manipulate the model's response by adding perturbations to conditional text or image, as shown in subfigure (a). The adversarial perturbation of the image can prevent diffusion model from editing the image. 2) Prompt Extraction attack shown in subfigure (b) aims to extract system prompt from the generated image. The high-quality prompts can be critical for some GenAI-based applications. 3) subfigure (c) illustrates Jailbreak attack where a diffusion model is misled to generate images with inappropriate content. The prompts with seemingly unharmful text can lead to inappropriate generation. 4) Backdoor attack in subfigure (d) manipulates training or fine-tuning process so that certain behavior can be induced by a pre-defined trigger without hurting normal usage. The presence of the symbol will add Donald Trump to the generated images since it is embeded during the training or finetuning process.}
        \label{fig:attacks_img}
\end{figure}
\subsection{Not To Generate Images for Harmful Instructions}
Recent advancements in T2I models enable various applications~\citep{ruiz2023dreambooth,gal2022image}. The power of the visual generative model also introduces a potential risk in the applications. Recent studies have revealed the vulnerability of T2I models to adversarial attacks, prompt extraction attacks, jailbreak attacks, and backdoor attacks. We now present the related work from these four types of attacks.

\subsubsection{Adversarial Attack on Text-to-Image Models}
\textbf{Attacking Text-to-Image Models for Bad.}
Recent studies examine the robustness of Diffusion models to variations in the input text, as shown in Fig.~\ref{fig:prom_inj_img}. The revealed models' low robustness has been leveraged to create attacks targeting specific image generation. An optimization-based approach is proposed to achieve target generation with subtle text prompts~\citep{liu2023riatig}. Additionally, the study suggests generating plausible text perturbations that humans might make, such as typos, glyphs, and phonetic variations~\citep{gao2023evaluating,du2024stable}. Both approaches require access to the models and their gradients, which may not always be feasible. In a black-box setting, an adversary can create adversarial prompts using an open-source text encoder~\citep{zhuang2023pilot}, although this encoder is still part of the Diffusion models. Furthermore, the research reveals hidden vocabularies in DALLE-2~\citep{daras2022discovering}, and make-up words can manipulate generation~\citep{milliere2022adversarial}. Based on the observations, a character-level optimization method based solely on text input is proposed to obtain adversarial prompts~\citep{kou2023character}.

A recent analysis thoroughly examines the robustness of diffusion models in both white-box and black-box settings~\citep{zhang2023robustness}. It investigates the robustness of each component of diffusion models and identifies the Resnet module in the decoder as highly vulnerable. Additionally, some adversaries aim to induce models to generate harmful images by circumventing prompt filters and safety mechanisms~\citep{yang2023mma,qu2023unsafe}, known as Jailbreak attack, which will be discussed later in this section.

\textbf{Attacking Text-to-Image Models for Good.}
Recent T2I models enable customized creation of visual content, which raises concerns about security and privacy, such as copyright infringement. Adversarial attacks on T2I models can also be applied to protect privacy by preventing editing based on T2I methods with adversarial noise on input images, as shown in Fig.~\ref{fig:prom_inj_img}. Preventing the creation of GAN-based deepfakes has been extensively researched~\citep{yeh2020disrupting,ruiz2020disrupting,huang2021initiative,wang2022anti,yang2021defending}. However, these methods cannot be easily applied to popular diffusion models due to several reasons~\citep{van2023anti}: 1) The generator of GAN is fixed, while the diffusion process of diffusion models is iterative and difficult to differentiate. 2) Text prompt information is integrated into each generation step of diffusion models, unlike GANs. 3) Some applications of diffusion models involve fine-tuning on a few-shot inputs, such as personalization in DreamBooth.

Several adversarial attacks tailored for diffusion models have been developed to combat copyright infringements. One approach is to manipulate the image feature representation to match a specific target in the latent space defined by the diffusion model's encoder~\citep{salman2023raising}. The work~\citep{shan2023glaze} proposes to apply style-transferred versions of the original image as viable targets in the latent space. The efficacy of these attacks is verified in the work where they show the latent space is the bottleneck to achieve high attack effectiveness~\citep{xue2023toward}.

The latent space-based attacks overlook the influence of textual prompts during the diffusion process, which can still leak significant information into the generated images. To address this, the study suggests taking text prompts into consideration and targeting the entire reconstruction loss directly~\citep{salman2023raising}. One challenge is the difficulty in obtaining gradients of the iterative diffusion process. This is mitigated by using only a few steps~\citep{salman2023raising} or obtaining expected gradients through Monte Carlo sampling~\citep{liang2023adversarial}. Additionally, the study discovers that reducing the number of time steps can enhance the effectiveness of adversarial noises~\citep{wang2023simac}. Therefore, they propose an adaptive greedy search method to find the optimal number of time steps.

Many current adversarial examples for diffusion models are tailored to specific models and don't transfer across different situations. To tackle this, the study in~\citet{liang2023mist} suggests combining the two types of attacks above to enhance the transferability of adversarial examples across various diffusion models and their applications. Additionally, the work~\citep{rhodes2023my} proposes using multiple losses together to enhance this transferability even further.

One application of diffusion models is personalization, which involves fine-tuning model parameters using just a few examples~\citep{ruiz2023dreambooth,gal2022image}. This poses new challenges for attacking diffusion models. To address this challenge, the study introduces Alternating Surrogate and Perturbation Learning~\citep{van2023anti}. In this approach, adversarial noise and model parameters are optimized alternately to achieve effective adversarial noises. Additionally, \citet{zhao2023unlearnable} formulate this problem as a max-min optimization problem and introduce a noise scheduler-based method to enhance the effectiveness of the adversarial attacks.

The robustness of protective noises created by adversarial attacks has been explored~\citep{liang2023mist,zhao2023can,salman2023raising,liang2023adversarial}. The study demonstrates that selecting a suitable target image can enhance the robustness of the noise against noise purification~\citep{liang2023mist}. Additionally, \citep{zhao2023can} introduce an effective purification technique capable of removing such protective noises.

Furthermore, the concept of generating protective noises against diffusion model-based personalization has been extended to video inputs. For example, building on previous work~\citep{salman2023raising}, the study in~\citet{li2024prime} presents an efficient method to generate protective noise for video generation models.

\subsubsection{Prompt Extraction Attack on Text-to-Image Models} 
Recent advancements in T2I generation models have opened up various applications, such as artwork design~\citep{cao2023difffashion,yang2024artfusion}. Creating high-quality prompts can be time-consuming and costly. Many efforts have been made to develop effective prompts~\citep{gu2023systematic,hao2024optimizing,liu2022design,oppenlaender2023taxonomy}. However, as illustrated in Fig.~\ref{fig:prom_ext_img}, recent research has unveiled the possibility of leaking prompts for image generation, known as the Prompt Extraction attack~\citep{shen2023prompt,leotta2023not}. One simple method for extracting text prompts involves using an image captioning model on a generated image. However, the prompts for high-quality images are often complex and cannot be easily achieved with standard caption models~\citep{li2022blip,mokady2021clipcap}. Another explored method for prompt extraction is optimization-based, where text is iteratively updated to achieve high semantic similarity with the generated image. The text with the highest similarity is taken as the extracted prompt~\citep{prompteng}. However, this approach is computationally expensive and requires many manually defined hyperparameters.

The research indicates that prompts for generating high-quality images should include a subject along with several prompt modifiers~\citep{liu2022design,oppenlaender2023taxonomy,gu2023systematic}. Inspired by these findings, \citet{shen2023prompt} propose a method that utilizes a caption model to capture the subject and a multi-label classifier to predict the prompt modifiers. The two outputs are then combined to generate the final stolen prompt. Furthermore, \citet{leotta2023not} delve into prompt extraction within a specific context: generating images with an artist's style. It makes the first exploration of how to identify an artist's name within the input string, given the generated image. In addition, defending against prompt extraction is crucial for protecting intellectual property. \citet{shen2023prompt} demonstrate that adding an optimized adversary noise to the generated images can disrupt effective extraction methods.

\subsubsection{Jailbreak Attack on Text-to-Image Models}
\textbf{Jailbreak Attack.} Recent Diffusion models like Stable Diffusion (SD) are trained on large-scale datasets containing image-text pairs from the web~\citep{rombach2022high}. There's a concern that these models might generate inappropriate images since the datasets also include harmful concepts. \citet{schramowski2023safe} demonstrate that SD indeed generates biased content, sometimes even reinforcing such biases. As shown in Fig.~\ref{fig:jail_img}, inappropriate degeneration occurs on a large scale across various text-to-image generative models, even with normal text prompts~\citep{brack2023mitigating}. Moreover, attackers in~\citet{yang2023mma} aim to alter textual prompts while maintaining their semantic intent, resulting in the generation of targeted NSFW (Not Safe For Work) content that may bypass existing filters. \citet{qu2023unsafe} investigate how adversaries create text prompts to generate specific types of unsafe content, such as widely disseminated hateful memes.

\textbf{Defense against Jailbreak Attack.}
To prevent Diffusion models from generating inappropriate content, various approaches have been explored. One intuitive method is to reject or alter prompts that might lead to unsafe outputs~\citep{brack2023mitigating,ni2023ores,gandikota2023erasing,kumari2023ablating,ni2023ores}. However, even seemingly harmless text prompts can sometimes result in inappropriate content, such as the prompt "a gorgeous woman" generating a nudity image. Similar concerns exist for defenses based on text embedding spaces~\citep{chuang2023debiasing,struppek2023rickrolling}, which are not always effective.

Another approach involves removing inappropriate content from the training data and retraining the model from scratch on the cleaned dataset~\citep{gandikota2023erasing,schramowski2023safe}. However, this method is computationally expensive and may not entirely prevent the generation of inappropriate content. To mitigate the cost of retraining, a proposed solution is to fine-tune diffusion models or text embeddings to unlearn harmful concepts and promote safer generations~\citep{gandikota2023erasing,kumari2023ablating,gandikota2024unified}. Recent research shows that harmful concepts are not fully removed by the popular unlearning methods~\citep{pham2023circumventing,tsai2023ring}. Additionally, methods have been developed to guide generation away from unsafe concepts without requiring fine-tuning~\citep{schramowski2023safe,brack2023mitigating}. This involves applying classifier-free guidance to steer generation away from harmful content. Furthermore, identifying directions in feature space corresponding to harmful concepts and modifying query activations accordingly can contribute to safer generation~\citep{li2023self}.

Post-hoc approaches have also been explored, where inappropriate images are detected using a safety guard classifier and rejected~\citep{gandhi2020scalable,birhane2021large,rando2022red}. However, the effectiveness of this approach largely depends on the performance of the detection classifier.

\textbf{Evaluation of Jailbreak Attack.}
A fair and comprehensive evaluation is crucial for advancing safe Diffusion models within the community. Evaluation of jailbreak performance typically involves two types of text prompts: 1) natural prompts like Inappropriate Image Prompts (I2P)~\citep{schramowski2023safe}, and 2) adversarial prompts intentionally crafted by adversaries to induce inappropriate generation~\citep{yang2023mma}. The performance of jailbreak is typically measured using the Area Under the Curve (AUC) score. Common classifiers such as the Q16 classifier~\citep{schramowski2022can} and NudeNet~\citep{yang2023mma} are employed to assign probabilities to generated images indicating their likelihood of being unsafe.

\subsubsection{Backdoor Attack on Text-to-Image Models}
Backdoor attacks can manipulate model behavior by inserting tainted samples into the training data or altering the training process with specific trigger patterns~\citep{gu2019badnets}. Traditionally, these attacks have been focused on classification tasks~\citep{gu2019badnets,li2022backdoor}. However, with the recent advancements in T2I models, researchers have begun exploring backdoor attacks on visual generative models. Unlike standard backdoor attacks on classifiers, those targeting Diffusion models aim for high utility and target specificity during inference. Essentially, the goal is for the backdoored T2I model to behave normally in the absence of a trigger but generate specific images upon receiving the implanted trigger signal, as illustrated in Fig.~\ref{fig:backdoor_img},.

Previous research in this area has investigated backdoor attacks on various generative models such as GANs~\citep{goodfellow2020generative} and VAEs~\citep{kingma2013auto}. For instance, during the inference stage, the generator of a GAN generates samples from noise sampled from a specified distribution. In a backdoored GAN scenario, the model is trained to generate normal samples from the typical prescribed sampling distribution while also producing targeted samples from a predefined malicious distribution~\citep{rawat2022devil}.

Recent research has started exploring backdoor attacks tailored for diffusion models. Specifically focusing on text-conditional diffusion models, the study in~\citet{struppek2023rickrolling} introduces backdoors by incorporating a backdoored text encoder. Examining the standard text-to-image pipeline, \citet{chou2023backdoor} further suggest backdooring various components involved in integrating conditional texts, such as the embedded tokenizer, the language model, and the U-Net architecture. Additionally, \citet{vice2023bagm} propose modifications to both the training data and the forward/backward diffusion steps to implant backdoor behaviors into unconditional diffusion models. Rather than creating model-specific backdoors, the work~\citep{chou2024villandiffusion} introduces a unified backdoor attack framework that can be applied to mainstream diffusion models with different schedulers, samplers, and conditional and unconditional designs.

Most backdoor attacks on diffusion models target specific images or images with particular attributes. Additionally, more fine-grained targets have been explored. The research introduces and analyzes three types of adversarial targets: instances belonging to a certain class from the in-domain distribution, out-of-domain distribution, and one specific instance~\citep{chen2023trojdiff}. Furthermore, the study explores three backdoor targets from a different angle, considering Pixel-Backdoor, Object-Backdoor, and Style-Backdoor~\citep{zhai2023text}. Backdoor attack triggers are often designed to be inconspicuous, with rare tokens frequently utilized~\citep{struppek2023rickrolling}. Moreover, the study demonstrates that common tokens used as triggers in benign text prompts can negatively impact image generation~\citep{zhai2023text}. The duration of backdoor behavior persistence has also been investigated. They reveal that the behavior gradually fades away during further training, suggesting potential for the development of backdoor defense methods for diffusion models~\citep{zhai2023text}. Besides, the detection and defense of backdoor attacks on diffusion models has also been explored~\citep{an2023remove,an2024elijah}.

An important application of diffusion models worth mentioning is personalization~\citep{ruiz2023dreambooth,gal2022image}. Personalization often aims to learn a new concept using only a few examples, sometimes just one. These new concepts can then be incorporated into image generation when a specific pattern is provided. Therefore, personalization can also be viewed as a form of backdoor. Due to its high computational efficiency and effectiveness with minimal examples, \citep{huang2023zero} suggest a personalization-based backdoor approach.

\subsection{Not To Generate Training Image}
\subsubsection{Membership Inference Attack on Text-to-Image Models.}
\begin{figure}[t]
     \centering
     \begin{subfigure}[b]{0.48\textwidth}
         \centering
         \includegraphics[width=0.9\textwidth]{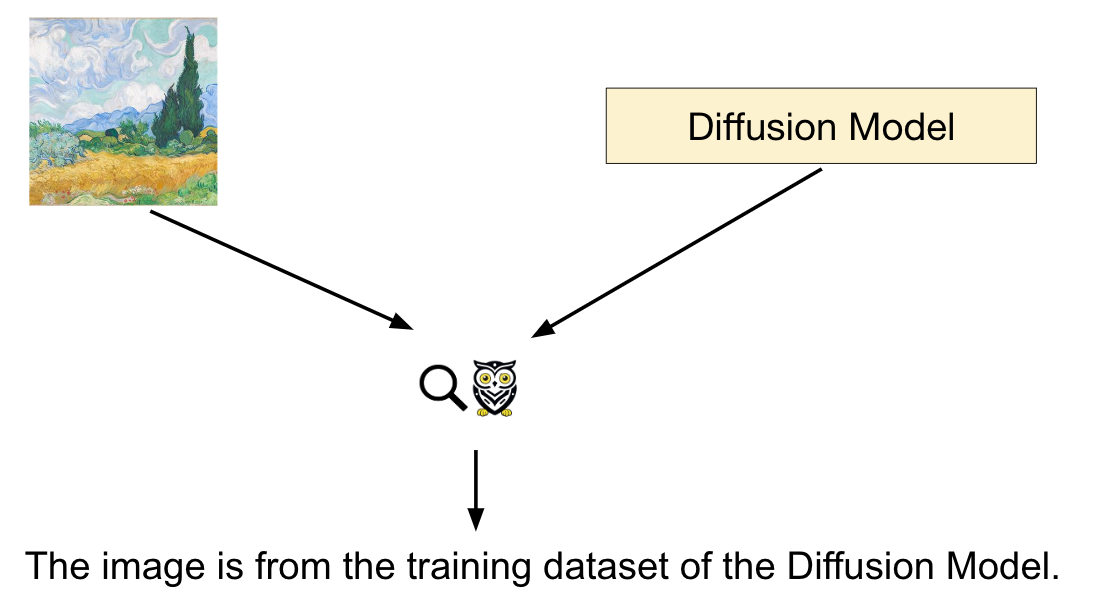}
         \caption{Image Membership Inference Attack}
         \label{fig:memb_infer_img}
     \end{subfigure}
     \hfill
     \begin{subfigure}[b]{0.38\textwidth}
         \centering
         \includegraphics[width=0.6\textwidth]{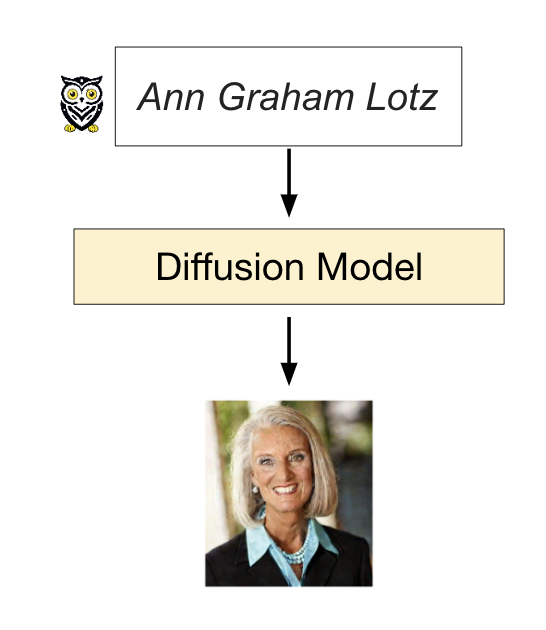}
         \caption{Training Image Extraction Attack}
         \label{fig:img_extract}
     \end{subfigure}
        \caption{Training data-related attacks on Text-to-Image models: Image Membership Inference attack aims to infer whether a particular image is from the training dataset. Moreover, Training Image Extraction attack shown in subfigure (b) aims to generate training images or objects in the image directly, e.g., the same identify as one from training images.}
        \label{fig:img_leak}
\end{figure}

Membership Inference Attack (MIA) aims to determine if a given sample originates from the training set of a model~\citep{shokri2017membership}, as shown in Fig.~\ref{fig:memb_infer_img}. MIA has been extensively studied in discriminative models~\citep{yeom2018privacy,salem2018ml,nasr2019comprehensive,choquette2021label}, relying on the behavioral differences between member and non-member samples. For instance, perturbations applied to member samples lead to larger prediction changes than those for non-members. Similarly, MIA has been explored for Diffusion models based on similar assumptions. The work~\citep{hayes2017logan} indicates that the logits of the discriminator from GANs can be applied to identify memberships effectively. Likewise, reconstruction loss can serve as an indicator for membership in VAE models~\citep{hilprecht2019monte}. Additionally, in cases where the target generative model is inaccessible, memberships can be identified by assessing the distance between synthetic samples and member samples. Synthetic samples generated by the target model tend to be closer to member samples than non-member ones~\citep{hu2023membership,chen2020gan,mukherjee2021privgan}.

Given the remarkable performance of diffusion models, there's been significant attention on MIA in these models. Studies indicate that existing MIAs designed for GANs or VAEs are largely ineffective on diffusion models due to various reasons~\citep{duan2023diffusion}, such as 1) inapplicable scenarios (e.g., requiring the discriminator of GANs) and 2) inappropriate assumptions (e.g., closer distances between synthetic samples and member samples). To address this, a new approach is proposed, which infers membership by comparing the loss values of member and test samples~\citep{hu2023membership}. Essentially, member samples are expected to have lower losses compared to non-member ones. To enhance attack effectiveness, a method called LiRA is introduced~\citep{carlini2022membership}. It involves training a set of shadow models on different subsets of the training set and computing their losses on test samples. The average losses of models containing the test samples in their training data are notably lower than those of the remaining models. Similarly, membership identification is also achieved by assessing the matching of forward process posterior estimation at each timestep, where member samples typically exhibit smaller estimation errors compared to hold-out non-member samples~\citep{duan2023diffusion}. A more efficient membership attack with two queries is further proposed in~\citet{kong2023efficient}. In addition to loss information, model gradients of diffusion models have also been utilized in MIAs~\citep{pang2023white}. Specifically, gradients from all diffusion steps are employed as features to train the attack model for MIA.

The hyperparameters of the diffusion model, such as timesteps, sampling steps, sampling variances, and text prompts, also influence the model's resistance against MIAs. The study in~\citep{matsumoto2023membership} indicates that timesteps play a significant role, with intermediate steps in the noise schedule being the most susceptible to attack. Additionally, sampling steps have a greater impact on MIA performance compared to sampling variances. Furthermore, information from text prompts can be directly utilized to identify membership based on the pairwise relationship between texts and corresponding images~\citep{wu2022membership}.

The study highlights that existing MIAs designed for GANs or VAEs are largely ineffective in diffusion models~\citep{duan2023diffusion}. However, it concludes that diffusion models exhibit comparable resistance to MIAs as GANs~\citep{matsumoto2023membership}. This apparent discrepancy in claims stems from differences in evaluation settings. Therefore, fair evaluation of MIAs is crucial. Common evaluation metrics include Attack Success Rate (ASR) and Area Under Receiver Operating Characteristic (AUC). To prioritize the importance of correctly inferring membership, Carlini argues for reporting True-Positive Rate (TPR) at an extremely low False-Positive Rate (FPR)~\citep{carlini2022membership}. Moreover, the choice of evaluation datasets is also critical. For instance, assuming that the member set and hold-out set come from different distributions can lead to reporting a very high ASR~\citep{wu2022membership}. However, MIA performance may be far from perfect when evaluating challenging datasets. Additionally, the composition of the training dataset can also impact MIA performance~\citep{golatkar2023training}. The study demonstrates that models trained on very small datasets with low internal variance show high resistance against MIAs, potentially overestimating model safety in real-world scenarios with diverse datasets.

\subsubsection{Training Data Extraction.}
The generation of training data using generative models poses significant threats to copyright protection. The reasons behind such replication have been investigated in the context of GANs. Studies have found that the replication tendency of GANs is inversely related to dataset complexity and size~\citep{feng2021gans}. Furthermore, GANs trained on face datasets not only produce replicated images but also generate novel images of identities from the training dataset~\citep{webster2021person}.

Recent research has also examined similar replication phenomena in diffusion models, as shown in Fig.~\ref{fig:img_extract}. \citet{somepalli2023diffusion} demonstrate that diffusion models can reproduce high-fidelity content from their training data. To address this, a generate-and-filter pipeline is proposed, enabling the extraction of over a thousand training examples from state-of-the-art models~\citep{carlini2023extracting}. Their results also reveal that diffusion models are more susceptible to training data extraction attacks compared to previous generative models like GANs. Building upon this, a more efficient extraction attack called template verbatim is proposed, significantly reducing network evaluations~\citep{webster2023reproducible}.

In terms of evaluating generative models, the Fréchet Inception Distance (FID) score is commonly used to assess the quality of generated images. However, FID tends to favor models that memorize training data~\citep{bai2021training}. To address this bias, the inclusion of authenticity scores is proposed, enabling the detection of noisy pixel-by-pixel copies during evaluation~\citep{alaa2022faithful}.

\textbf{Defense Against Memorization of Text-to-Image Models.}
The success of MIA can be attributed to the tendency of Diffusion models to memorize training samples during the training process. Various techniques have been explored to enhance model robustness against MIAs. One straightforward approach is to remove duplicate samples from the training set, as many popular datasets contain numerous duplicated samples~\citep{carlini2023extracting,dedupmiti}. However, even in the absence of duplicates, models can still memorize portions of the training data. Overfitting during training exacerbates this memorization. Therefore, improving model robustness can also involve reducing overfitting~\citep{fu2023probabilistic}. Another strategy to prevent training data leakage is to train multiple diffusion models on separate subsets of the data and then ensemble them during inference~\citep{golatkar2023training}.

The methods mentioned above offer empirical reductions in memorization but don't guarantee robustness against MIAs. As a theoretically grounded approach, differential privacy has also been explored in diffusion models. However, training Diffusion Models (DMs) using Differential Privacy Stochastic Gradient Descent (DP-SGD) significantly compromises generation quality~\citep{lyu2023differentially}. To address this challenge, a solution is proposed: pre-training DMs with public data, followed by fine-tuning them with private data using DP-SGD for a brief period~\citep{harder2022pre,ghalebikesabi2023differentially}. Additionally, training DMs with Differential Privacy (DP) is improved by adopting Latent Diffusion Models (LDMs), where only attention modules are tuned with privacy-sensitive data, significantly reducing computational costs~\citep{lyu2023differentially}. Furthermore, the study emphasizes the importance of DM parameterization and sampling algorithms in applying differential privacy. A modification of DP-SGD for DM training is proposed, further enhancing model robustness~\citep{dockhorn2022differentially}.

\begin{figure}[t]
     \centering
     \begin{subfigure}[b]{0.45\textwidth}
         \centering
         \includegraphics[width=0.81\textwidth]{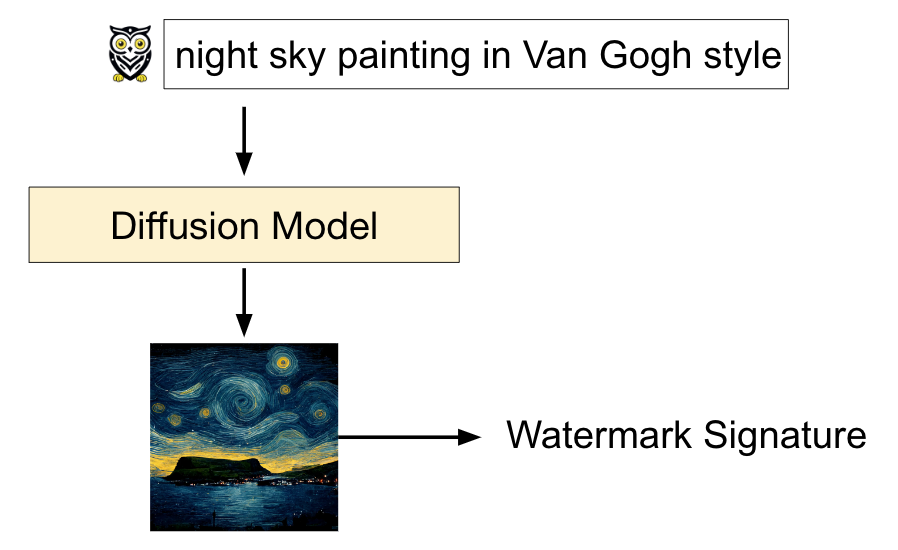}
         \caption{Watermarking Image Generation}
         \label{fig:watm_img}
     \end{subfigure}
     \hfill
     \begin{subfigure}[b]{0.45\textwidth}
         \centering
         \includegraphics[width=0.76\textwidth]{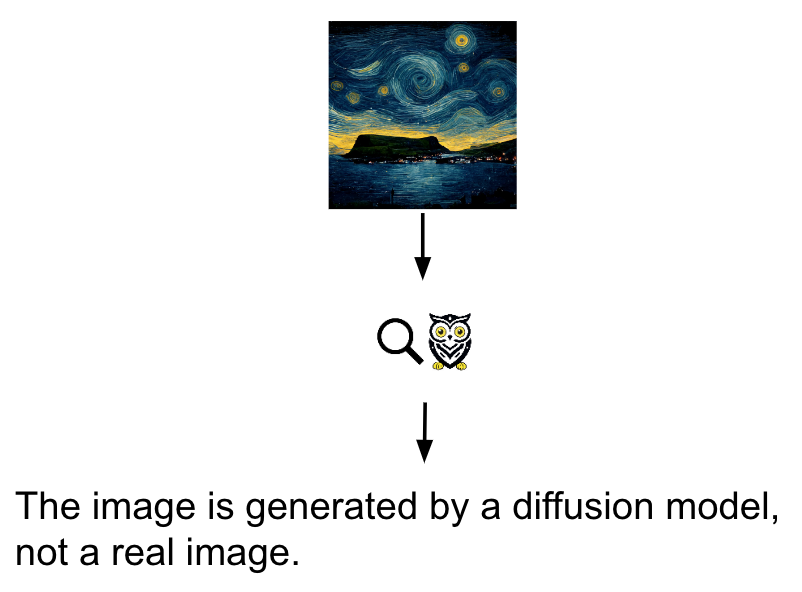}
         \caption{AI-generated Image Detection}
         \label{fig:img_det}
     \end{subfigure}
     
     \begin{subfigure}[b]{0.45\textwidth}
         \centering \vspace{0.5cm}
         \includegraphics[width=0.98\textwidth]{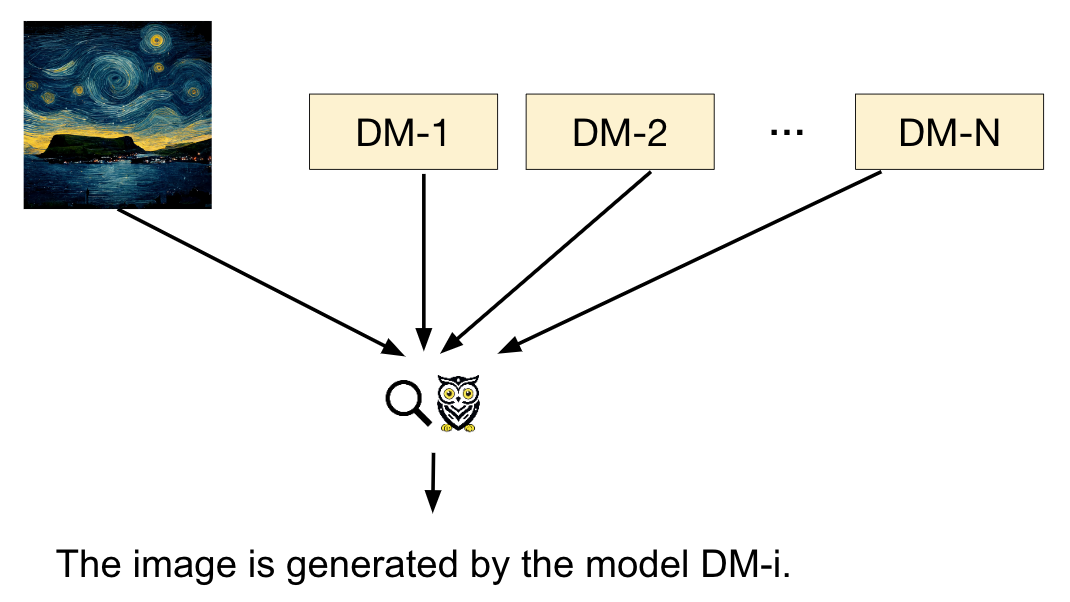}
         \caption{AI-generated Image Attribution}
         \label{fig:model_attr}
     \end{subfigure}
     \hfill
     \begin{subfigure}[b]{0.45\textwidth}
         \centering \vspace{0.5cm}
         \includegraphics[width=0.95\textwidth]{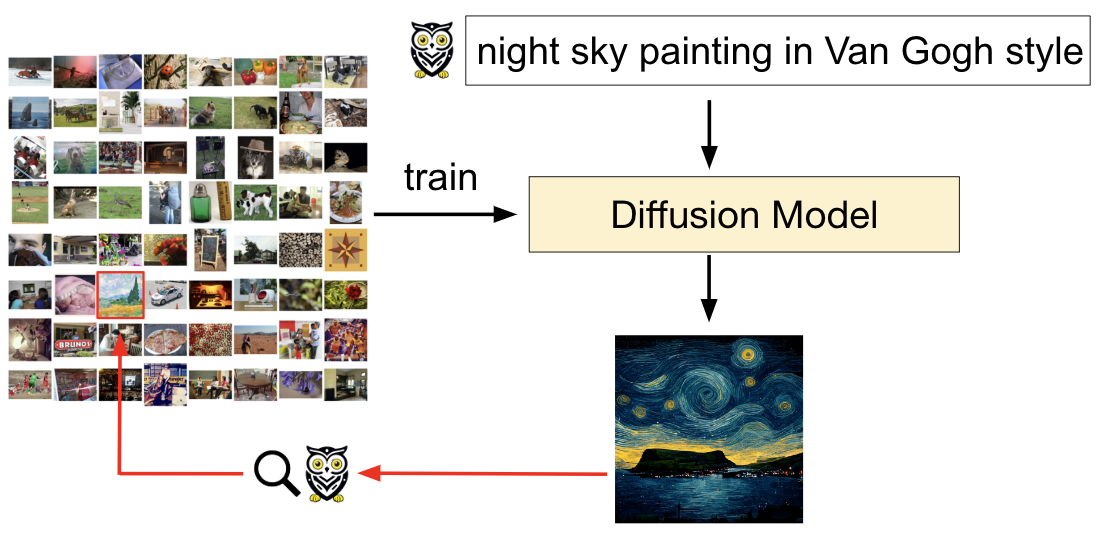}
         \caption{Data Attribution of Generated Image}
         \label{fig:data_attr}
     \end{subfigure}
        \caption{Identifiable Generated Image: Subfigure (a) shows a way to watermark generated images so that they can be identified later. Image Detection shown in subfigure (b) aims to distinguish the generated images from real ones, while Model Attribution in subfigure (c) aims to infer whether an image is generated by a given model. Data Attribution aims to find supporting training images for a generated image, shown in subfigure (d).}
        \label{fig:ident_img}
\end{figure}
\subsection{To Generate Identifiable Images}
As recent generative models produce highly realistic visual content (e.g. images and videos), often indistinguishable from real-world scenes, there's a growing need to identify the source of generated text to prevent potential misuse and protect intellectual property. To tackle these challenges, various watermarking techniques have been proposed for textual generation as a proactive measure~\citep{cui2023diffusionshield,wen2023tree,fernandez2023stable}. In cases where no watermark is available, research has also focused on the detection and attribution of generated visual content. Detection aims to distinguish between generated and real images~\citep{coccomini2023detecting,yu2021survey}, while attribution aims to identify the generative model responsible for a given image~\citep{sha2023fake,kim2020decentralized}. Furthermore, studies on data attribution have aimed to identify which training images are relevant to a generated image, further enhancing intellectual property protection~\citep{park2023trak,ilyas2022datamodels}.

\subsubsection{Watermarking of Generated Image.}
Watermarking images illustrated in in Fig.~\ref{fig:watm_img} has a long history, involving methods to embed imperceptible information into images for later extraction and ownership verification~\citep{o1996watermarking,cox1996secure}. With the advent of modern deep neural networks, there are new opportunities and challenges for enhancing image watermarking techniques~\citep{hayes2017generating,zhu2018hidden,liu2022watermark}. However, applying these methods to generated images directly can impact their quality. Additionally, standalone watermarking stages can be easily removed or disregarded when generative models are made open-sourced, such as Stable Diffusion~\citep{rombach2022high}. Consequently, researchers have begun studying watermarking techniques integrated into the generation process itself.

One straightforward approach is to train or fine-tune generative models on images with pre-defined watermarks, ensuring that all generated images are watermarked~\citep{yu2021artificial,zhao2023recipe,cui2023diffusionshield}. Furthermore, watermark information can be embedded into generative models from latent space, such as latent dimensions in GANs~\citep{yu2020responsible,nie2023attributing} and initial noise in Diffusion models~\citep{wen2023tree}. However, intervening in the entire generation process is computationally intensive. To address this, it is possible to selectively fine-tune only the decoder of generative models~\citep{fei2022supervised,fernandez2023stable}, compelling it to generate watermarked images more efficiently.

An indirect method to establish ownership of generated images is by claiming ownership of the generative model responsible for their creation. Previous watermarking techniques have mainly targeted discriminative models. They can be broadly categorized into two groups~\citep{peng2023intellectual}: static watermarking~\citep{uchida2017embedding,wang2021riga}, which embeds a specific pattern in the static content of the model, such as model parameters, and dynamic watermarking~\citep{adi2018turning,zhang2018protecting,li2019prove}, which embeds a similar pattern in the model's dynamic contents, such as its behavior.

With recent advancements, watermarking generative models have garnered significant attention. Some approaches propose watermarking Generative Adversarial Networks (GANs) by establishing mappings between trigger inputs and outputs provided by the generator, using regularization constraints~\citep{yu2021artificial,fei2022supervised}. However, these techniques cannot be straightforwardly applied to diffusion models due to their markedly different data modeling approaches. Nevertheless, recent studies have put forth watermarking methods tailored for diffusion models. For instance, one approach involves fine-tuning the diffusion model on images containing watermarks, ensuring that generated images also carry embedded watermarks~\citep{yu2021artificial,zhao2023recipe,cui2023diffusionshield}. Explorations have also been made into embedding watermarks based on conditions for image generation, where specific patterns presented in the condition lead the model to generate corresponding images~\citep{liu2023watermarking}. However, these approaches are limited to conditional generative models. To address these limitations, a study introduces a watermark diffusion process that requires neither modification of training nor condition input for generation~\citep{peng2023protecting}.

\subsubsection{Detection of AI-generated Image.}
As a passive measure, AI-generated image detection aims to differentiate fake images from real ones, as shown in Fig.~\ref{fig:img_det}. Existing Detection approaches can be summarized into two groups: 1) The first group works by analyzing the forensic properties of generated images, such as semantic inconsistencies (e.g., irregular eye reflections)~\citep{hu2021exposing}, known generation artifacts in the spatial~\citep{nataraj2019detecting}, and artifacts in the frequency domain~\citep{frank2020leveraging}. 2) The second group uses neural networks to learn a feature space where representations of fake and real images can be distinguished~\citep{wang2020cnn}.

Numerous methods have been proposed for detecting GAN-based images, particularly focusing on deepfakes~\citep{mirsky2021creation}. With the rise of Diffusion models (DM), attention has shifted to detecting generic-generated images. Generalizing GAN detection approaches to DMs is a natural step. However, existing detectors trained on GAN images struggle to distinguish real from DM-generated ones~\citep{ricker2022towards,corvi2023detection}. Retraining these detectors on DM-generated data significantly improves their performance~\citep{ricker2022towards,corvi2023detection}. New solutions for detecting DM-generated images have emerged, exploring lighting and perspective inconsistencies~\citep{farid2022lighting,farid2022perspective}. DMs often produce physically implausible scenes, which can be detected by the difference in reconstruction accuracy compared to real images~\citep{wang2023dire}.

Given the rapid evolution of generative models, it is crucial to develop detectors that can generalize to new generators. While detectors designed for GAN-generated images struggle with DM-generated ones, the reverse surprisingly works well~\citep{coccomini2023detecting}. The argument is that DM-generated images have fewer detectable artifacts, making them more challenging to identify than GAN-generated ones. One possible reason for this is the absence of grid-like frequency artifacts, a known weakness of GANs, in DM-generated images~\citep{ricker2022towards}. Additionally, efforts are underway to create universal detection methods applicable across different generative models. Specifically, \citet{ojha2023towards} suggest using a pre-trained vision transformer with a classification layer, instead of a classifier based on fake and real images.

Text associated with images has also been explored in detection. When available, image-related text can enhance detection performance~\citep{coccomini2023detecting,sha2023fake}. For real images, these texts might be captions, while for generated images, they could be prompts used during generation. A method involves building an MLP classifier using features extracted by both a CLIP vision encoder and a text encoder~\citep{coccomini2023detecting}.

The community has also delved into Generated Video Detection, particularly focusing on deepfake videos~\citep{yu2021survey}. An intuitive approach involves identifying visual anomalies in the video, such as boundary irregularities~\citep{li2020face,li2018exposing}, abnormal biological signals~\citep{li2018ictu,ciftci2020fakecatcher}, and consistency issues characterized by camera fingerprints~\citep{lukas2006digital,cozzolino2019noiseprint}. This approach often requires domain expertise to extract relevant features. Conversely, a straightforward end-to-end approach detects fake videos by treating the video as a sequence of images and applying fake image detectors to each frame. Many image detectors with various network architectures have been proposed for this purpose, including traditional classification models~\citep{zhou2017two,rossler2019faceforensics++,deng2022deepfake} and manually designed novel alternatives~\citep{afchar2018mesonet,nguyen2019capsule,deng2022deepfake}. Additionally, a temporal-consistency-based approach has been explored, utilizing networks with sequential modeling capabilities~\citep{guera2018deepfake,montserrat2020deepfakes,masi2020two}. Temporal consistency can also aid in data processing for detection, such as computing the optical flow of the video~\citep{amerini2019deepfake}. However, most of these approaches have not been validated on generated generic videos. With recent advancements in video generation, such as SoRA~\footnote{https://openai.com/sora}, there is still much to explore in detecting these generated generic videos.

\subsubsection{Model Attribution of AI-generated Image.}
As illustrated in Fig.~\ref{fig:model_attr}, model attribution of generated images aims to solve the following problem: which generative models generate a particular image? Studying this question can aid in identifying and holding responsible users behind the misuse of such images.

Researchers have conducted model attribution based on the principle that a synthetic sample is best reconstructed by the generator that created it~\citep{wang2023alteration, laszkiewicz2023single}. This process typically requires access to the parameters or gradients of the target generative models. Additionally, model-agnostic attribution methods have been explored. For example, one approach involves training a multi-class classifier as an attribution like the approaches developed for GAN~\citep{liu2024model,bui2022repmix,yu2019attributing,marra2019gans}. Recent research proposes to conduct model attribution with only a few shot samples from target models~\citep{liu2024model}. Researchers further investigated how prompts used to generate fake images influence both detection and attribution~\citep{sha2023fake}. They discovered that fake images can be accurately attributed to their source models by identifying unique fingerprints within the generated images. Moreover, they found that prompts related to certain topics, like \textit{"person"}, or with a specific length, between 25 and 75, facilitate the generation of more authentic fake images. However, relying on a centralized classifier is not scalable, as it necessitates retraining when new generative models are introduced~\citep{kim2020decentralized}. Instead, the proposed solution involves decentralized attribution, where a binary classifier is constructed for each model. Each binary classifier is then used to differentiate images generated by its associated model from those generated by others.

\subsubsection{Data Attribution of AI-generated Image.}
In contrast to model attribution, data attribution of generated images seeks to identify which images in the training set have the greatest impact on the appearance of a given generated image, as shown in Fig.~\ref{fig:data_attr}.

One traditional method for implementing data attribution on machine-learning models is the influence function~\citep{koh2017understanding}. This method estimates the effect of removing a data point from the training set by approximating the resulting parameters through Taylor expansion. However, it cannot be trivially applied to diffusion models for two main reasons: it is not scalable to deep models with large training datasets~\citep{feldman2020neural}, and it is unreliable in non-convex settings~\citep{basu2020influence}. Another commonly used approach is ensemble-based, where many models trained on subsets of the entire training dataset are examined. A recent study~\citep{dai2023training} has applied the ensemble-based approach to diffusion models, but they only conducted analysis on small-sized generated images. Scaling the ensemble-based data attribution approach to large-scale training datasets is challenging. To address this issue, a solution is proposed, which conducting image retrieval in a pre-defined feature space~\citep{wang2023evaluating}. This method assumes that synthesized images are influenced by training images that are close to them in the defined feature space. For example, the feature space could be provided by CLIP encoders. However, the attribution performance of this approach is sensitive to the defined feature space.

To achieve a balance between effectiveness and efficiency, the researchers propose TRAK (Tracing with the Randomly-projected After Kernel)~\citep{park2023trak} and extend it to diffusion models~\citep{georgiev2023journey}. The work~\citep{zheng2023intriguing} conducts empirical studies on data attribution with diffusion models and observes that design choices for attribution, though theoretically unjustified, can empirically outperform previous baselines significantly.

To better assess data attribution for Diffusion models, the researchers suggest a method to identify the ground truth training images that influenced a synthesized image~\citep{wang2023evaluating}. They achieve this by taking a pre-trained generative model and fine-tuning it on a new exemplar image. As a result, the images generated by the tuned model are computationally influenced by the exemplar.

Several evaluation metrics have been proposed to quantitatively assess data attribution, two of which are computationally tractable for diffusion models. One metric involves counterfactual evaluation~\citep{ilyas2022datamodels}, which calculates the pixel-wise L2-distance and CLIP cosine similarity of images generated by models trained with or without the exclusion of the most relevant images identified by an attribution method. Another metric proposed in the study is called the linear data modeling score~\citep{park2023trak}, which measures the model's ability to accurately predict counterfactual outcomes when the training set is modified in a specific manner.

\section{Responsible Generative AI in Safety-critical Applications}
\label{sec:apps}
In this section, we delve into the application of responsible generative AI across various domains, including healthcare, education, finance, and artificial general intelligence. Specifically, we highlight the risks and concerns stemming from the limitations of current generative AI, with a primary focus on technical aspects.

\subsection{Responsible Generative AI for Healthcare}
Both textual and visual generative models have diverse applications in the healthcare sector~\citep{shokrollahi2023comprehensive}. Visual generative models, such as diffusion models~\citep{rombach2022high}, are extensively utilized in medical imaging tasks like medical image reconstruction~\citep{gungor2023adaptive,xie2022measurement}, medical image-to-image translation~\citep{lyu2022conversion,ozbey2023unsupervised}, medical image generation~\citep{pan20232d,muller2023multimodal}, medical image classification~\citep{oh2023diffmix,yang2023diffmic}, medical image registration~\citep{kim2022diffusemorph}, and medical image segmentation~\citep{kim2022diffusion,azad2022transdeeplab}. On the other hand, textual generative models, like transformer-based LLMs~\citep{openai2023gpt4}, find applications in protein structure prediction~\citep{behjati2022protein,castro2022transformer,boadu2023combining}, clinical documentation and information extraction~\citep{sivarajkumar2022healthprompt,yogarajan2021transformers}, diagnostic assistance~\citep{azizi2022enhanced,zhou2023transformer}, medical imaging and radiology interpretation~\citep{chaudhari2022application,nimalsiri2023automated}, clinical decision support~\citep{meng2021bidirectional,wang2023acquisition}, medical coding and billing~\citep{lopez2023explainable,ng2023modelling}, as well as drug design and molecular representation~\citep{bagal2021molgpt,li2022kpgt}. The effectiveness of ChatGPT~\citep{openai2023gpt4} and DALL-E~\citep{betker2023improving} in some of these applications is examined, and the strengths and limitations of healthcare-customized LLMs like Med-PaLM~\citep{singhal2023large} and BioGPT~\citep{luo2022biogpt} are compared and discussed in~\citep{sai2024generative}.

The integration of generative AI into the healthcare sector has received significant attention, accompanied by various efforts. However, numerous risks and concerns have emerged during this integration~\citep{kuzlu2023rise}. Some from a technical perspective are as follows.

\begin{itemize}
    \item Demand for large-scale training data with sensitive information: Collecting medical data, often containing sensitive information, poses challenges due to privacy concerns. Generative AI models require extensive training data for optimal performance~\citep{bandi2023power,jadon2023leveraging}.
    \item Consequences of failed decisions: Generative AI, including GAI, may yield unreliable results due to limited generalization abilities in real-world scenarios~\citep{huang2023survey}. Incorrect decisions made by AI models concerning patient data can have severe consequences, including harm or even threat to life, which is unacceptable.
    \item Lack of interpretability: Decision-making by generative AI in healthcare necessitates explanations~\citep{bharadiya2023generative,dunn2023generative}. While current models can offer textual rationales for their predictions, these explanations may not accurately reflect their decision-making process~\citep{rajani2019explain,huang2023can,zhao2024explainability}.
    \item Bias and discrimination: Training data for AI models in healthcare may exhibit biases, leading to biased outcomes favoring specific groups~\citep{sap2019social}. Detecting and mitigating such biases in generative AI is challenging.
    \item Medical data privacy: Risks of medical data leakage exist at various stages, including data collection, training, and model deployment~\citep{chen2024generative}. Recent research suggests that training data can even be extracted directly from LLMs~\citep{carlini2019secret,carlini2021extracting}.
\end{itemize}   

Addressing these limitations requires advancements in generative AI itself, alongside the development of articulated regulations for real-world applications~\citep{varghese2023chatgpt}.

\subsection{Responsible Generative AI for Finance}
Recent advancements in textual generative AI, such as ChatGPT~\citep{openai2023gpt4}, offer enhanced capabilities in understanding text, which finds applications in finance. These applications can be grouped into three main areas~\citep{chen2023fiction}: providing customized services, risk management, and decision support.

Firstly, GenAI facilitates automated customer service, leading to improved efficiency, cost reduction, and enhanced customer experience~\citep{chen2023fiction,dahal2023utilizing}. For instance, financial institutions can leverage GenAI to comprehend customer needs, engage directly with customers, and tailor marketing strategies accordingly. Secondly, GenAI enables risk analysis with natural language explanations~\citep{wang2023generative,chen2023fiction}. For example, lenders can utilize GenAI to assess loan requests and receive guidance on whether to lend to a particular borrower. Lastly, GenAI supports management and decision-making processes~\citep{chen2023fiction,dahal2023utilizing}. For instance, individual investors lacking professional analysis skills can utilize GenAI to identify reliable investment opportunities. In addition to these applications, GenAI has been employed to address various other challenges in finance, such as generating financial data~\citep{assefa2020generating,naritomi2020data,eckerli2021generative}, which is out of our discussion.

The limitations of GenAI present risks and concerns for its applications in the financial sector~\citep{remolina2023generative,shabsigh2023generative,rane2023role}. Common concerns include:
\begin{itemize}
    \item Financial hallucinations: GenAI may produce inaccurate or nonsensical outputs, potentially impacting risk assessment processes and risk management negatively~\citep{roychowdhury2024journey,huang2023survey}.
    \item Explainability in financial decision-making: Understanding the decision-making process of GenAI is challenging due to its complex network architecture. While GenAI provides textual explanations for its decisions, these explanations may not accurately reflect the decision process~\citep{rajani2019explain,huang2023can,zhao2024explainability}.
    \item Fairness in financial decision-making: Biases in the training data and input prompts of GenAI can result in discriminatory outcomes or perpetuate societal inequalities~\citep{sap2019social}.
    \item Financial data protection: GenAI's ability to generate training data poses a risk of data leakage when financial data is used for training or fine-tuning~\citep{carlini2019secret,carlini2021extracting}.
    \item Systemic risk and financial stability: Automation of real-time decisions and the unreliability of decision-making tools may contribute to systemic risk in the financial sector.
    \item Fraud detection in finance: GenAI can be exploited by fraudsters to impersonate customer service representatives, leading to fraudulent activities that are difficult to detect~\citep{ahmadi2023open}.
    \item Cybersecurity risks: GenAI-generated content may be exploited for malicious purposes~\citep{galle2021unsupervised,abburi2023simple}, and new cyberattacks targeting large-scale GenAI systems may emerge, such as energy attacks that disrupt GenAI services~\citep{shumailov2021sponge,gao2024inducing}.
\end{itemize}

Recent research has conducted a comparative analysis of various generative models in financial applications. The study~\citep{krause2023large} delves into the performance of different large-language models within the finance sector. Furthermore, \cite{rane2024gemini} examines and contrasts Gemini and ChatGPT in depth. The findings indicate that Gemini, benefiting from Google's extensive knowledge base and search capabilities, excels in accuracy and depth. On the other hand, ChatGPT demonstrates creativity and proficiency in text generation, making it adept at producing concise summaries and engaging in conversational interactions. Additionally, there have been proposals for generative models tailored specifically to the financial domain, such as Bloomberg GPT~\citep{bloomberg2023introducing} and Morgan Stanley's GPT-4 variant~\citep{davenport2023morgan}. As with other fields, addressing the risks and concerns necessitates collaboration among experts from various backgrounds, not solely technical contributors.

\subsection{Responsible Generative AI for Education}
Recent advancements in textual generative models have paved the way for their application in education. Particularly, conversation-based models like ChatGPT~\citep{openai2023gpt4} have garnered widespread attention across various sectors. The advantages of leveraging these models, such as ChatGPT, are multifold~\citep{baidoo2023education}. Firstly, they facilitate personalized tutoring by offering tailored guidance and feedback to individual students based on their unique learning requirements and progress. Secondly, they enable interactive learning experiences by engaging in conversational interactions that take into account contextual history. Thirdly, they can automate the evaluation of essays, allowing educators to allocate their time more efficiently to other teaching tasks. Lastly, these models can provide real-time feedback and assessment, offering insights within seconds and reducing the workload for instructors.

However, there are also some risks and concerns associated with the technical limitations of generative AI. These can be summarized as follows:
\begin{itemize}
    \item Generating biased learning materials: Since generative models are trained on biased data, they may produce biased content~\citep{sap2019social}.
    \item Misuse of learning tools: Generative models can be used to provide learning content but can also be exploited to generate harmful or inappropriate content. Recent research~\citep{zou2023universal} has shown that adversarial prompts can manipulate models like ChatGPT to follow harmful instructions, even if they are aligned not to do so. Additionally, these tools can also be misused for plagiarism~\citep{maronikolakis2020transformers,abburi2023simple}.
    \item Generating wrong or non-factual content: Textual generative models, such as large language models (LLMs), are known to likely generate content that is inconsistent with inputs or even contradicts factual information~\citep{huang2023survey}.
    \item Lack of creativity: Generative models rely on statistical patterns from their training data and may provide feedback that lacks creativity, even for students requiring innovative suggestions.
    \item Limited performance in certain disciplines: Generative models may not perform well in certain tasks, for example, the ones that involve complex mathematical computation or deep reasoning~\citep{frieder2024mathematical,liu2023evaluating}.
\end{itemize}

In addition to the limitations caused by technical issues, other concerns also exist. For instance, GenAI-based learning systems lack human interaction, which may be less effective for students who prefer personal connections with teachers. Moreover, the quality and accessibility of AI-driven learning tools significantly impact learning performance in different regions.

Moreover, visual generative models have been explored in education~\citep{vartiainen2023using,han2023design,dehouche2023s}. For instance, they can be used to teach art history, aesthetics, and techniques. Learning systems based on visual generative models face similar risks and concerns as those based on textual generative models. Especially, one important question to address is the ownership of artistic works during the learning and teaching process~\citep{liu2023watermarking}.

To promote the use of AI in education, the responsibilities of AI should be further studied. Additionally, policymakers, researchers, educators, and technology experts should collaborate, as GenAI-based education systems involve various perspectives. 

\subsection{Responsible Generative AI for Artificial General Intelligence}
Broadly defined, Artificial General Intelligence (AGI) refers to machines that can perform any
intellectual task that a human being can do~\citep{goertzel2007human}. Technically speaking, while Artificial Narrow Intelligence (ANI) corresponds to AI models tailored to specialized tasks (e.g. image recognition~\citep{krizhevsky2012imagenet}, chess playing~\citep{silver2016mastering}, or video generation~\citep{sora}), AGI aims to combine these various skills into a single system demonstrating general intelligence. Currently, two roadmaps can be observed to extend generative AI to AGI~\citep{zhang2023one}, namely, coordination strategy and unified strategy.

The coordination strategy takes a textual generative model (e.g. LLM~\citep{openai2023gpt4}) as a central controller that analyzes the main task, assigns subtasks to other agents, and coordinates their outputs to get the final decision. For example, a task for AGI is 'Please introduce the history of the German national flag with an illustration of texts, images, and videos'. The controller will decompose the task into subtasks, assign them to corresponding agents (generative models for text, image, and video), and summarise all the outputs together in a logical order as final outputs. Many efforts have been made in this direction~\citep{suris2023vipergpt,su2023pandagpt,li2023multimodal,wu2023next}.

Another possible strategy is the unified strategy that builds a powerful model solving all tasks within the model~\citep{openai2023gpt4,peng2023kosmos,driess2023palm}. For example, for a given 'Please introduce the history of the German national flag with an illustration of texts, images, and videos', the model can generate the multimodal story directly without the help of any other models. For example, GPT-4V~\citep{openai2023gpt4} can respond to both text understanding and image understanding tasks. More capability can be integrated into the models, e.g., visual and audio generation. 

It is worth mentioning that Embodied AI plays a crucial role in the development of AGI by bridging the gap between abstract reasoning and real-world interaction. Unlike traditional AI systems that operate solely in virtual environments, embodied AI integrates physical embodiment, allowing AGI to perceive and interact with the world much like humans do. This embodiment enables AGI to gather sensory information, understand context, and learn from physical experiences, leading to more robust and adaptable intelligence~\citep{duan2022survey}. In current literature, generative models are applied to understand and interact with the real world in embodied AI.

Generative models have made significant strides in the development of AGI, but they also pose challenges that impact their integration into AGI systems.

\begin{itemize}
    \item Wrong decisions caused by the hallucination of generative models: Generative models are known to hallucinate unexisting concepts and may produce erroneous outputs, leading to wrong decisions by AGI~\citep{bang2023multitask,barrett2023identifying}.

    \item Bias and source of bias: Generative models can inherit biases present in their training data, potentially perpetuating societal biases in AGI behavior. The biased behaviors of AGI can be caused by multiple factors, which are hard to identify and remove~\citep{sap2019social,barrett2023identifying}.

    \item Attacks on perception of AGI: Adversarial attacks can exploit vulnerabilities in generative models to deceive AGI perception systems by misleading perception modules. The understanding and behavior based on wrong perception is unexpected and worrying~\citep{}.

    \item Data privacy of AGI systems: Generative models trained on sensitive data may inadvertently leak private information through generated outputs, posing privacy risks in AGI applications~\citep{carlini2023extracting}.

    \item Copyright of AGI systems: Generated content produced by AGI systems raises questions about copyright ownership and intellectual property rights. Establishing legal frameworks and ethical guidelines for copyright attribution and ownership of AI-generated works is essential to address this issue.

    \item Decomposition of task and Coordination of subtasks: Generative models can facilitate task decomposition by generating diverse solutions for complex problems, enabling AGI to break down tasks into manageable subtasks. The decomposition and coordination can potentially be manipulated to mislead the final AGI behaviors~\citep{khot2022decomposed,wang2024stop}.

    \item Coordination of multi-agent: Generative models can support the coordination of multiple agents in AGI systems by generating coherent and collaborative behaviors. Coordination mechanisms, such as communication protocols and negotiation strategies, might be vulnerable to perception-based manipulation, which can induce more unexpected behavior of AGI.
\end{itemize}

In conclusion, while generative models offer substantial benefits for AGI development, addressing their limitations is crucial to ensure the robustness, fairness, privacy, and effectiveness of AGI systems in real-world applications.

\section{Challenges and Oppotunities}
\label{sec:dis}

\textbf{Developing Robust and Efficient Harmful Content Detection Methods}: Generative models can generate toxic content, following harmful instructions or even non-toxic instructions. A practical way to address this is to always detect generated toxic outputs with detectors, such as toxicity classifier~\citep{dathathri2019plug}, Q16 classifier~\citep{schramowski2022can} and NudeNet~\citep{yang2023mma}. However, these toxic content detectors are neither robust to adversarial input nor generalizable to new toxic contents. Thus, creating robust mechanisms to detect toxic content generated by generative models is critical. Challenges include accurately identifying various forms of harmful content, including misinformation, hate speech, and graphic imagery~\citep{touvron2023llama}, across different modalities such as text and images. Additionally, there is a need to balance detection accuracy with computational efficiency to enable real-time monitoring and response. Opportunities lie in building efficient models to identify harmful content with high precision and recall. Overall, the robustness and the efficiency of harmful content detectors remain to improve.

\textbf{Aligning Text-to-Image Models with Human Values}: Harmful content detectors show limited detection performance and require extra computational cost for each inference. To address that, it is important to align the generative model with human values, e.g., avoiding the generation of toxic content. For instance, post-training is conducted to align pre-trained LLMs with human values, as introduced in Sec.~\ref{sec:text_pre}. Similar post-training has also been applied to text-to-image models so that they can better follow users' instructions~\citep{lee2023aligning,wu2023human}. However, they mainly focus on the quality of the generated images regarding the text prompts and largely overlook the toxicity of generated images. Ensuring that text-to-image models generate non-toxic content is also important. Import future work could be conducting alignment from the responsible perspective and developing evaluation metrics and frameworks that assess the alignment of generated images with human values, e.g., fairness, bias, and toxicity.

\textbf{Adapting to Evolving Human Value and User Preference}: Ethical considerations surrounding generative AI can evolve in response to societal values, cultural norms, and technological advancements~\citep{cobarrubias1983ethical}. The evolvement also brings challenges to the alignment of current generative models, since the current alignment approach assumes a fixed human value. Overcoming the challenges requires models to be dynamically aligned with the evolving ethical standards and norms. A similar challenge also exists with user preference. For example, the same instruction specified by the same user can indicate different things at different times. Hence, it is important to develop adaptable approaches that enable generative models to incorporate evolving ethical guidelines and norms into their decision-making processes.

\textbf{Enhancing User Control and Transparency}: Harmful instructions might induce generative models to generate toxic content. A response user can specify the responsible requirement in the prompts. However, existing generative models are not guaranteed to follow the responsible instructions either. Empowering users with greater control over the generated content and fostering transparency in the generative process is essential for responsible usage of GenAI. In addition to the input prompts, more intuitive interfaces should be designed and integrated so that users can specify preferences and constraints on the generated outputs.

\textbf{Exposing Vulnerabilities in Embodied AI Systems}: Generative models (i.e. LLMs), as an important central controller, have been applied to embodied AI systems that interact with the physical world through sensors, actuators, and a central controller~\citep{brohan2022rt,brohan2023rt}. The system can inherit the vulnerability of LLMs. When exploited in embodied AI, it can face more security challenges due to its physical presence and potential for real-world impact. Concretely, adversaries could manipulate or disrupt system behavior by identifying vulnerabilities in perception, decision-making, action execution modules, and their fusion modules. Research remains to be conducted to uncover vulnerabilities and assess their potential impact on safety and security. By revealing vulnerabilities in embodied AI, researchers can better develop robust systems against potential threats.

\textbf{Studying the Risks Brought by Generative Models in More Domains}: Investigating the potential risks and unintended consequences of deploying generative models across diverse domains is crucial. In this paper, we briefly discuss the potential risks and concerns brought by generative models in healthcare, education, finance, and artificial general intelligence in Sec.~\ref{sec:apps}. We believe that there are many more domains where GenAI can be broadly applied, e.g., cybersecurity, environmental science, and urban planning. Hence, it is critical to discuss, reveal, and mitigate the concerns brought by generative models before they are deployed in real-world applications.

\section{Conclusion}
\label{sec:conclu}

Generative AI has emerged as a powerful tool with applications across various domains, from natural language processing to image generation. However, as generative models increasingly find applications in real-world scenarios, it is critical to ensure that the generated content is not only high-quality but also responsible. In this survey, we have highlighted the responsible requirements of current generative models, focusing on two main categories: textual and visual generative models. We provided a unified perspective on the responsibility of both textual and visual generative models and identified five key practical responsible requirements, namely, truthfulness, impartiality, safety, data privacy, and copyright clarity. These requirements address fundamental concerns associated with generated content. 

Our discussion regarding the risks and concerns associated with the application of GenAI in real-world scenarios calls for the attention of the community. Furthermore, we discuss the challenges and opportunities for responsible GenAI, which can inspire more research. Besides, it is imperative for researchers, practitioners, and policymakers to collaborate closely to develop robust frameworks, tools, and guidelines for ensuring the development of responsible GenAI and the responsible use of GenAI. We hope this paper can benefit the community in this direction.

\noindent\textbf{Acknowledgement:} I would like to thank Google Responsible AI team for the feedbacks, especially Dr. Ahmad Beirami and Dr. Kathy Meier-Hellstern. I would also like to express my thank to the safety teams of Torr Vision Group at University of Oxford and Tresp Lab at University of Munich. This work is supported by the UKRI grant: Turing AI Fellowship EP/W002981/1, EPSRC/MURI grant: EP/N019474/1.

\bibliography{main}
\bibliographystyle{plainnat}

\end{document}